\documentclass[a4paper,12pt]{article}
\usepackage[margin=2cm]{geometry}
\usepackage{graphicx}
\usepackage{xcolor}
\usepackage{multicol}

\usepackage{amsmath,amsthm,amssymb,mathabx}
\allowdisplaybreaks[3]
\usepackage{mathdots,dsfont}
\usepackage{enumitem}
\newcommand{\innerproduct}{non-degenerate symmetric bilinear form}

\theoremstyle{definition}
\newtheorem{prop}{Proposition}

\newtheorem{lem}{Lemma}
\newtheorem{ex}{Example}
\newtheorem{defn}{Definition}

\newtheorem{rmk}{Remark}
\newtheorem*{warn*}{Warning}

\newcommand{\step}[1]{\vskip5pt\noindent\textsc{#1}}

\usepackage{datetime2}

\usepackage[T1]{fontenc} 
\usepackage{libertinus}
\DeclareRobustCommand{\textsc}[1]{%
	{\fontfamily{LibertinusSerif-TLF}\selectfont\scshape #1}%
}

\usepackage[backend=bibtex,sorting=ynt,giveninits=true,doi=false,isbn=false,url=false]{biblatex}
\usepackage[
  colorlinks=true,
  linkcolor=blue,
  citecolor=green!50!black,
  urlcolor=red!50!black
]{hyperref}

\title{Second-order superintegrable systems from semi-simple and nilpotent Frobenius structures}
\author{Andreas Vollmer\\[0pt] \small University of Hamburg, Dept.\ of Mathematics, \small Bundesstr.~55, 20146 Hamburg, Germany\\[-2pt] \small\texttt{andreas.vollmer@uni-hamburg.de}}
\date{\today}

\begin{document}
\maketitle

\begin{abstract}
    Recently, it was shown that a rich class of second-order (maximally) superintegrable systems has an underpinning Hesse-Frobenius structure, i.e.\ a Frobenius structure that is compatible with a Hessian structure such that the Hessian pre-potential is also a Frobenius pre-potential.
    Hence, these superintegrable systems arise, locally, from (possibly non-unital) Frobenius algebras.
    
    We present a conification to lift systems of non-zero constant sectional curvature to flat ones, thus reducing the problem on the Frobenius algebra side to the associative case. Furthermore, we employ a direct product construction to generate higher-dimensional second-order maximally superintegrable systems on flat spaces.
    
    We apply this approach to some basic semi-simple and nilpotent algebras and explicitly construct the arising second-order superintegrable systems using a Neumann series expansion.
    We find that all non-degenerate second-order maximally superintegrable systems in three dimensions arise from these examples.
    
\end{abstract}

\section{Introduction}\label{sec:introduction}

Second-order superintegrable systems are Hamiltonian systems with a maximal amount of functionally independent constants of the motion (aka \emph{first integrals} or simply \emph{integrals}). They play a pivotal role in the Sciences. Key examples are the Harmonic oscillator and the Kepler-Coulomb system, which provide crucial models in classical and quantum mechanics as well as celestial mechanics. Second-order superintegrable systems also arise naturally in mathematics, e.g.\ in the study of surface metrics that admit infinitesimal symmetries of their unparametrised geodesics \cite{Lie1882,Matveev2012,BMM2008,MV2020}.
Second-order superintegrable systems have been intensively studied from many angles, such as via separation of variables \cite{KKM2018_book,BM2017}, geodesic equivalence \cite{Vollmer2020,MV2019,Valent2016}, coalgebras \cite{MZZ2023,BBHMR2009}, via limiting procedures \cite{CKP2015,KMP2013,KKMP2007} and warped product geometries \cite{CDR2015,CR2017,CR2024}. Furthermore, they have been linked to orthogonal polynomials organised in the Askey-Wilson scheme \cite{KMP2013,KMP2007,KMP2011}.

A classification of second-order (maximally) superintegrable (Hamiltonian) systems exists to date in dimension two, and, partially, in dimension three, see \cite{KKM2005_I,KKM2005_II,KKM2005_III,KKM2006_IV,KKM2006_V,KKPM2001,Nikitin2022_1,Nikitin2023_2,Nikitin2023_3,Nikitin2024_4,Nikitin2025_5}.
A classification in both of these dimensions exists, in particular, for \emph{non-degenerate} second-order superintegrable systems, which admit a $(n+2)$-parameter potential.
Higher dimensional systems, in contrast, are still much less understood. Some families that exist in arbitrary dimension are known, such as the Smorodinski-Winternitz systems  and the so-called \emph{generic system on the $n$-sphere} \cite{KSV2024,CHMZ2021,KKMP2007,Iliev2018}. The latter takes its name from the fact that in dimensions two and three, all other non-degenerate second-order superintegrable systems can be obtained from it via Stäckel transformations (i.e.\ special conformal rescalings of the system) and via contractions \cite{KKM2005_I,KKM2005_II,KKM2005_III,KKM2006_IV,KKMP2007}. It has been conjectured that this is true also for higher dimensions.
\medskip

This paper builds on recent results that facilitate an effective and efficient study of second-order superintegrable systems in higher dimensions.
Note that the investigation of the systems of partial differential equations that underlie second-order superintegrability becomes increasingly cumbersome and extensive as the dimension progresses. Therefore, the pre-existing methods for their study rapidly grew intractable.
This made the classification problem virtually unmanageable for higher dimensions.
In response to this challenge, algebraic geometry had been proposed as a promising method to classify second-order superintegrable systems in arbitrary dimension \cite{MPW2013,KKM2007_3Dflat,KKM2007_2Dflat,KS2018}. A new geometric approach along this direction has recently been developed, see \cite{KSV2023,KSV2024,KSV2024_2D}. Crucially, it remains manageable in arbitrarily high dimension. Indeed, it has led to a geometric characterisation, in arbitrarily high dimension, of non-degenerate systems for which all integrability conditions are satisfied generically, see \cite{Vollmer2025_Frobenius,Vollmer_HF,CV2025}; such systems are called \emph{abundant systems}. This characterisation has several advantageous properties: it is tensorial (i.e.\ it provides a geometric description independent of a choice of coordinates) and it naturally incorporates Stäckel transformations (as conformal rescalings, or more precisely in terms of a Weyl structure), for instance \cite{Vollmer2025_Frobenius,Vollmer2025_Weyl,KSV2023,KSV2024}.
It also provides an algebraic and information-geometric characterisation of these systems \cite{AV2025,CV2025,Vollmer2025_Frobenius,Vollmer_HF}.

In particular, abundant systems whose underlying metric is of constant sectional curvature and dimension\footnote{In dimension two the correspondence is not 1-to-1, as there are superintegrable systems that do not arise from Hesse-Frobenius structures, cf.\ \cite{CV2025}.} $n\geq3$ are shown to correspond 1-to-1 to curved Frobenius structures with a compatible Hessian structure (we later call them \emph{Hesse-Frobenius structures}), cf.\ \cite{KSV2023,Vollmer2025_Frobenius,Vollmer_HF}. This means, in particular, that the metric underlying such systems can be written locally in the form
\begin{equation*}
	g_{ij} = \partial_i\partial_j\phi
\end{equation*}
for a locally defined function $\phi$, where $\partial_i$ denote derivatives with respect to a flat structure. More precisely, the superintegrable system is encoded in a flat torsion-free connection $\mathtt D$, and the $\partial_i=\frac{\partial}{\partial x^i}$ arise from a choice of $\mathtt D$-affine coordinates $(x^1,\dots,x^n)$.
\medskip

The purpose of this paper is to use the underpinning Hesse-Frobenius structure to systematically construct examples of abundant (non-degenerate irreducible second-order maximally superintegrable) systems in any dimension.
The foundation for this approach was laid elsewhere, e.g.\  \cite{KKM2018_book,KSV2023,KSV2024,KSV2024_2D,Vollmer2025_Frobenius,Vollmer_HF}.
For the purposes here, it is important to highlight that abundant systems on spaces of constant sectional curvature ($n\geq3$) arise, locally around a point on the manifold, as solutions of a system of partial differential equations whose initial data are (possibly non-unital) Frobenius algebras. This renders the classification problem entirely algebraic and, conceptually speaking, it allows us, for $n\geq3$, to switch between three distinct viewpoints:
\begin{enumerate}[label=(\roman*)]
	\item abundant systems on spaces of constant sectional curvature,
	\item Hesse-Frobenius structures (i.e.\ differential-geometric Frobenius structures),
	\item the \emph{initial data} of such a structure, which is a commutative and (non-)associative algebra with a (Frobenius-)compatible \innerproduct.
\end{enumerate}
A novelty of this paper is that the recently developed framework of \cite{KSV2023,Vollmer2025_Frobenius,Vollmer_HF} is employed here to \emph{explicitly} construct local solutions in a \emph{systematic} manner, rather than just stating their existence, making the general framework tangible and concrete. To facilitate this, we invoke additional techniques that are only available through the correspondence of abundant systems and Hesse-Frobenius structures. In particular, we use a Neumann series expansion, which provides an effective means for constructing the local second-order maximally superintegrable system from its algebraic initial data.

As a second innovation, this paper puts forth two techniques that allow one to obtain new Hesse-Frobenius structures from lower-dimensional examples. These simplify the study and classification of abundant systems, and they highlight the power of the underlying framework. Indeed, the two techniques, while being quite easy on the side of Frobenius structures, are non-obvious constructions on the superintegrable side. One of these techniques is a conification, which implies a significant simplification on the algebraic side as it removes non-associativity from the underlying Hesse-Frobenius structure.
The second technique is a direct product construction, with which new examples can be produced by joining known lower-dimensional ones. This construction exhibits a remarkable property from the viewpoint of superintegrability: the Killing tensor fields associated to the constituent superintegrable systems, which survive on the direct product, typically account only for a small fraction of all the Killing tensor fields associated to the product superintegrable system. The existence of most of the associated Killing tensor fields is enforced and encoded by the Hesse-Frobenius structure.

As a concrete application, we utilise these techniques and construct second-order superintegrable systems from a few basic, explicit algebras as initial data.
Somewhat astonishingly, these examples already account for all local examples of flat three-dimensional non-degenerate second-order irreducible superintegrable Hamiltonian systems.
This sheds new light on their classification. We find the underlying (commutative and associative) Frobenius algebras and thus provide a novel way of thinking about these systems and their interrelations. Compared to the existing literature, the resulting algebraic characterisation turns out to be strikingly concise.
\bigskip

The paper is organised as follows:
in Section~\ref{sec:method}, we first recap some necessary prerequisites.
Section~\ref{sec:HF2SIS} then presents the procedure for constructing abundant systems from Hesse-Frobenius structures on spaces of constant sectional curvature, based on the framework established by \cite{KSV2023,Vollmer2025_Frobenius,Vollmer_HF}.
We then proceed to the algebraic viewpoint, initially limiting to the case when $g$ is flat, which corresponds to associativity on the Hesse-Frobenius side. In Section~\ref{sec:sis.from.algebra}, we explain how \emph{flat} superintegrable systems can be obtained by integration of the potentiality condition \eqref{eq:HF:potentiality} for an initial Frobenius algebra.
We then discuss how flat Hesse-Frobenius structures on direct products can be crafted from lower-dimensional flat Hesse-Frobenius structures, see Section~\ref{sec:direct.products}.
The first part of the paper is concluded in Section~\ref{sec:conification}, where we readmit curvature: we present a conification that lifts Hesse-Frobenius structures with curvature to flat ones. On the algebraic side, this reduces the non-associative initial data problem to the simpler associative one.

The second part of the paper is devoted to examples. In Section~\ref{sec:semisimple}, we begin with semi-simple Frobenius algebras, including their degenerations. A specific example is given in Section~\ref{sec:3+1}.
We then proceed to various nilpotent examples, see Section~\ref{sec:nilpotent}. In Section~\ref{sec:SW.and.sphere}, we explain how the Smorodinski-Winternitz and the generic superintegrable system on the $n$-sphere are linked via conification.

Abundant systems are characterised by the existence of a large number of associated Killing tensor fields. It is therefore instructive to inspect how these arise in a direct product: in Sections~\ref{sec:2+2} and~\ref{sec:8D} we consider a 4D and an 8D product system, respectively, and illustrate how the construction ensures the existence of non-obvious Killing tensor fields in addition to those directly inherited from the constituent superintegrable systems.

We close the paper in Section~\ref{sec:discussion} with a discussion, comparing our results to the pre-existing classification of flat three-dimensional non-degenerate second-order maximally superintegrable systems (over $\mathbb C$), leading to a strikingly concise algebraic characterisation of these systems.

\section{Method}\label{sec:method}

Let $g\in\Gamma(\mathrm{Sym}_2(T^*M))$ be a non-degenerate symmetric $(0,2)$-tensor on a simply connected and oriented manifold $M$ of constant sectional curvature $\kappa$ ($g$ will be called a \emph{metric} in the following). We work over a field $\mathbb K$, which can be either $\mathbb R$ or $\mathbb C$.
In the case $\mathbb K=\mathbb R$, any signature is allowed, i.e.\ $g$ can be Riemannian or pseudo-Riemannian.
We consider the natural Hamiltonian of the form
\begin{equation}\label{eq:Hamilton}
H(x,p)=g^{ij}(x)p_ip_j+V(x)
\end{equation}
(Einstein's summation convention applies), where $(x,p)=(x^1,\dots,x^n,p_1,\dots,p_n)$ are canonical Darboux coordinates and where $V$ is a potential that depends only on the position coordinates, i.e.\ on $(x^1,\dots,x^n)$ only.

An integral of the motion for $H$ is a function $F(x,p)$ that Poisson commutes with $H$, i.e.
\begin{equation}\label{eq:poisson}
	\{F,H\} = \sum_{j=1}^n \left( \frac{\partial F}{\partial x^j}\frac{\partial H}{\partial p_j} - \frac{\partial H}{\partial x^j}\frac{\partial F}{\partial p_j} \right) = 0.
\end{equation}
Integrals of the motion are constant along solutions of Hamilton's classical equations of motion.

For the next definition, we recall that $N$ functions $F^{(k)}:T^*M\to\mathbb K$, $0\leq k\leq N$, are said to be \emph{functionally independent} if $dF^{(0)}\wedge\dots\wedge dF^{(N)}\ne0$ on an open and dense subset of $T^*M$. Consequently, in order to have a dynamical system, $N=2n-2$ is the maximum possible.

\begin{defn}\label{defn:superintegrable.system}
A (maximally) \emph{superintegrable system} for the Hamiltonian $H$ is defined by $2n-2$ additional integrals of the motion $F^{(1)},\dots,F^{(2n-2)}$, such that $(H,F^{(1)},\dots,F^{(2n-2)})$ are functionally independent.
A (maximally) superintegrable system is \emph{second-order} if these integrals of the motion can be chosen to be of the form
\begin{equation}\label{eq:integral}
	F(x,p)=K^{ij}(x)p_ip_j+W(x)\,.
\end{equation}
\end{defn}

Without loss of generality, we assume that $K^{ij}=K^{ji}$.
It is straightforward to show that the coefficient functions $K^{ij}$ are associated to a Killing tensor field with components $K_{ij}$, where indices are lowered using the metric $g$.
Indeed, for \eqref{eq:integral}, the condition \eqref{eq:poisson} is a cubic polynomial in the fiber coordinates $p_i$ (called \emph{momenta}) with two homogeneous polynomial parts, namely a cubic and a linear one. The Killing condition,
\begin{equation}\label{eq:Killing}
	\nabla_kK_{ij}+\nabla_iK_{jk}+\nabla_jK_{ki}=0,
\end{equation}
follows from the cubic homogeneous part.
The linear homogeneous part, on the other hand, implies	
\begin{equation}\label{eq:preBD}
	\partial_jW = K_{ja}g^{ab}\,\partial_bV = {K_{j}}^{a}\partial_aV\,.
\end{equation}
Assuming $K$ and $V$ are known, 
\eqref{eq:preBD} can be integrated for $W$, if its integrability condition is satisfied. To compute it, we take a further derivative of both sides in~\eqref{eq:preBD}, and then skew symmetrise. This leads to the so-called \emph{Bertrand-Darboux condition} \cite{Darboux1901,Bertrand1857}
\begin{equation}\label{eq:bertrand-darboux}
	0 = (\partial_k\partial_j-\partial_j\partial_k)W 
	= \partial_k{K_{j}}^{a}\,\partial_aV-\partial_j{K_{k}}^{a}\,\partial_aV
	+{K_{j}}^{a}\partial_k\partial_aV-{K_{k}}^{a}\partial_j\partial_aV\,.
\end{equation}
The Bertrand-Darboux condition \eqref{eq:bertrand-darboux} can equivalently be cast into the concise form
\begin{equation}\label{eq:dKdV}
	d(K(\mathrm{grad}_gV)) = 0.
\end{equation}
Killing tensor fields $K$ that satisfy~\eqref{eq:dKdV} for a Hamiltonian \eqref{eq:Hamilton} can be extended, via \eqref{eq:preBD}, to an integral~\eqref{eq:integral}. We therefore call them the \emph{associated Killing tensor fields}.
\bigskip

Frobenius algebras and Hesse-Frobenius structures are the other two crucial definitions for the ensuing discussion.
\begin{defn}
	A \emph{Frobenius algebra} is a (finite-dimensional) commutative and associative algebra $\mathcal A$ over a field $\mathbb K$ furnished with a \innerproduct\ $\rho:\mathcal A\times\mathcal A\to\mathbb K$ such that the Frobenius property $\rho(ab,c)=\rho(a,bc)$ is satisfied. We say that it is \emph{unital} if there is an element~$\mathsf{u}$ such that $\mathsf{u}a=a$ for all $a\in\mathcal A$.
\end{defn}
In the following, the multiplication in a Frobenius algebra is usually denoted by $\blackdiamond$. We remark that a \innerproduct\ with the Frobenius property is sometimes also called a \emph{Frobenius form} or a \emph{Frobenius inner product}.

\begin{warn*}
We alert that the symbol $1$ will not always stand for the unit element $\mathsf{u}$, such as in Section~\ref{sec:nilpotent} below.
\end{warn*}

We now introduce Hesse-Frobenius structures. These may be viewed as the differential-geometric counterpart to commutative (non-)associative algebras equipped with a Frobenius inner product and whose associator is controlled by a constant.
In particular, adopting this viewpoint, flat Hesse-Frobenius structures are the counterpart to Frobenius algebras.

\begin{defn}
	A \emph{Hesse-Frobenius structure} on the manifold $(M,g)$ of constant sectional curvature $\kappa$ is a commutative product $\star:\mathfrak X(M)\times\mathfrak X(M)\to\mathfrak X(M)$ of vector fields, such that
	\begin{align}
		g(X\star Y,Z) &= g(X,Y\star Z)
		~\text{(Frobenius property)}
		\label{eq:HF:frobenius}
		\\
		[\hat C(X),\hat C(Y)]Z &= \kappa\left( g(X,Z)Y-g(Y,Z)X \right)
		~\text{(associator property)}
		\label{eq:HF:associator}
		\\
		(\nabla_Z\star)(X,Y) &= Z\star(X\star Y)
						+\kappa\left(2g(X,Y)Z+g(X,Z)Y+g(Y,Z)X)\right)
		\label{eq:HF:potentiality}
	\end{align}
	hold for all $X,Y,Z\in\mathfrak X(M)$, where $\nabla$ is the Levi-Civita connection of $g$ and where $\hat C(X)=X\star$ is an endomorphism on $\mathfrak X(M)$.
	We call a Hesse-Frobenius structure \emph{unital} if there is a vector field $\mathsf{e}$ such that $\mathsf{e}\star X=X$ for any vector field $X$.
	The triple $(M,g,\star)$ for a Hesse-Frobenius structure $\star$ on $(M,g)$ is going to be called a \emph{Hesse-Frobenius manifold}.
\end{defn}

Note that the underlying metric $g$ of a Hesse-Frobenius manifold is always of constant sectional curvature $\kappa$.
Furthermore note that the condition~\eqref{eq:HF:associator} is indeed a condition on the associator of $\star$, since
\[
	\mathrm{Assoc}(X,Y,Z)
	= (X\star Y)\star Z-X\star(Y\star Z)
	= [\hat C_Z,\hat C_X]Y.
\]
The product $\star$ hence need not be associative, and indeed it is only associative if $\kappa=0$, this is if $g$ is flat.
For the upcoming discussion, it is useful to introduce the cubic tensor field $C\in\Gamma(\mathrm{Sym}_3(T^*M))$ defined via
\begin{equation}\label{eq:defn.C}
	C(X,Y,Z)=g(X\star Y,Z),
\end{equation}
i.e.\ $X\star Y=X^iY^jC_{ijk}g^{kl}\partial_l$.
Note that, due to commutativity and the Frobenius property~\eqref{eq:HF:frobenius}, $C$ is totally symmetric.

\begin{rmk}
	To prevent misunderstandings, we mention that Hesse-Frobenius manifolds are special cases of curved Frobenius manifolds \cite{KKLNS2017,KKLNS2018,Mokhov2008,Vollmer_HF}. In the flat case ($\kappa=0$), they generalise Frobenius manifolds as introduced by Manin \cite{Manin1999}. However, note that Manin's definition is more general than Dubrovin's, which imposes some additional requirements \cite{Dubrovin1996,Dubrovin1998}. Most notably, here we do not require the existence of a unit or of an Euler vector field.
	On another note, the potentiality condition \eqref{eq:HF:potentiality} required here is more restrictive than the one imposed by Manin, Dubrovin and Hertling, for instance, cf.\ \cite{Manin1999,Dubrovin1996,Hertling2002_moduli}. Indeed, \eqref{eq:HF:associator} ensures that the metric $g$ underlying a Hesse-Frobenius structure is Hessian with respect to the flat connections $\nabla\pm\star$. The potentiality property \eqref{eq:HF:potentiality} ensures that the Hessian pre-potential $\phi$ with respect to $\nabla-\star$ is simultaneously a Frobenius pre-potential. Indeed, introducing $C(X,Y,Z)=g(X\star Y,Z)$, we have $C(X,Y,Z)=(\nabla^3\phi)(X,Y,Z)+\kappa(2g(X,Y)(\nabla_Z\phi)+g(X,Z)(\nabla_Y\phi)+g(Y,Z)(\nabla_X\phi))$ \cite{Vollmer_HF}.
\end{rmk}

As discussed, a Hesse-Frobenius structure is associative precisely if the underlying metric $g$ is flat. In this case, the conditions of a Hesse-Frobenius manifold simplify:
\begin{itemize}[left=5pt]
	\item The tensor defined by $C_{ijk}$ is symmetric in all three indices.
	\item The potentiality property \eqref{eq:HF:potentiality} becomes
	\begin{equation}\label{eq:potentiality.flat}
		\nabla_lC_{ijk}=C_{ija}g^{ab}C_{klb}.
	\end{equation}
	\item The associator property \eqref{eq:HF:associator} becomes the \emph{Witten-Dijkgraaf-Verlinde-Verlinde} associativity equation
	\begin{equation}\label{eq:WDVV}
		g^{ab}\left(C_{ija}C_{klb}-C_{ika}C_{jlb}\right)=0.
	\end{equation}
	Due to \eqref{eq:potentiality.flat}, it can be written as the differential equation
	\begin{equation}\label{eq:WDVV.PDE}
		\sum_{a,b=1}^n g^{ab}\left(( \partial_i\partial_j\partial_a\phi)\, (\partial_k\partial_l\partial_b\phi)
		-( \partial_i\partial_k\partial_a\phi)\, (\partial_j\partial_l\partial_b\phi)\right)=0
	\end{equation}
	in a $\nabla$-affine coordinate system. Indeed, $C_{ijk}=\partial_i\partial_j\partial_k\phi$ due to \eqref{eq:potentiality.flat}, cf.~\cite{Ferus1981,Vollmer_HF,KSV2023,Vollmer2025_Frobenius,AV2025}.
\end{itemize}

We return to the side of second-order superintegrable systems.
Note that maximal superintegrability, according to Definition~\ref{defn:superintegrable.system}, requires $2n-1$ \textit{functionally} independent integrals of the motion, including the Hamiltonian.
However, if instead we only require \textit{linear} independence, further integrals may exist.
For example, this is the case for abundant systems: they admit a $(n+2)$-parameter potential and $\frac{n(n+1)}{2}$ linearly independent associated Killing tensor fields.

A second-order superintegrable system is said to be \emph{irreducible}, if its associated Killing tensor fields do not have a non-trivial invariant subspace in common, viewing them as endomorphisms via $g$.
Below we will use the following facts about irreducible second-order superintegrable systems from \cite{KSV2023,Vollmer2025_Frobenius,Vollmer_HF}:

\paragraph{Structure tensor:}
It was shown in \cite{KSV2023} that the potential of an irreducible second-order superintegrable system satisfies
\begin{equation}\label{eq:Wilczynski}
	\nabla^2_{ij}V=\hat T_{ij}^k\partial_kV+\frac1n\,g_{ij}\,\Delta V\,,
\end{equation}
where $\hat T_{ij}^k$ is a $(1,2)$-tensor field that depends on the space spanned by the associated Killing tensors.
An irreducible second-order superintegrable system is then called \emph{non-degenerate} if the integrability condition is generically satisfied.
A non-degenerate (irreducible) second-order superintegrable system is encoded in a $(1,2)$-tensor field $\hat T_{ij}^k$ which can be decomposed according to
\begin{equation}\label{eq:structure.tensor.decomposed}
	T_{ijk}=\hat T_{ij}^ag_{ak}=S_{ijk}+\bar t_ig_{jk}+\bar t_jg_{ik}+\bar t_kg_{ij},
\end{equation}
where $S_{ijk}$ is totally symmetric and where $\bar t_i$ are components of a closed $1$-form \cite[sect.~3]{KSV2023}.
For a non-degenerate second-order superintegrable system, the associated Killing tensor fields are solutions of
\begin{equation}\label{eq:shortcut}
	\nabla_kK_{ij} = \frac13\left( \breve{T}^a_{ji}K_{ak}-\breve{T}^a_{ki}K_{aj} + \breve{T}^a_{ij}K_{ak}-\breve{T}^a_{kj}K_{ai} \right),
\end{equation}
where $\breve{T}^k_{ji}=g^{ka}\hat T^b_{aj}g_{bi}$, see \cite[sect.~5]{KSV2023}.
\emph{Abundant systems} are then defined as non-degenerate second-order superintegrable systems such that the integrability conditions for \eqref{eq:Wilczynski} and \eqref{eq:shortcut} are satisfied generically \cite{KSV2023}.

\paragraph{Integrability conditions:} The integrability conditions for an abundant system can be interpreted in terms of (non-degenerate relative) affine hypersurface theory \cite{CV2025}. On simply connected manifolds $(M,g)$ of constant sectional curvature $\kappa$ and dimension $n\geq3$, the structure tensors of abundant systems are precisely those such that
\[
	C_{ijk} = \frac13\left( T_{ijk}+\frac{1}{n-1}g_{ij}\hat T_{ka}^a \right)
\]
defines a Hesse-Frobenius structure on $(M,g)$ via \eqref{eq:defn.C}, see \cite{KSV2023,Vollmer2025_Frobenius,Vollmer_HF}.
In particular, the pair $(V,K)$ of solutions to \eqref{eq:Wilczynski} and~\eqref{eq:shortcut} then satisfies \eqref{eq:bertrand-darboux}.
This establishes a \emph{local correspondence} between the data of a Hesse-Frobenius manifold and abundant systems in $n\geq3$.
\begin{warn*}
	We alert that, in dimension two, abundant (non-degenerate irreducible second-order maximally superintegrable) systems exist that do not correspond to Hesse-Frobenius structures, see~\cite{CV2025}.
\end{warn*}

\paragraph{Initial data:} The potentiality condition \eqref{eq:HF:potentiality} is a prolongation system (sometimes called a closed system) of first order. Its integrability condition is \eqref{eq:HF:associator} \cite{KSV2023}.
By repeated substitution of $\nabla C$ via \eqref{eq:HF:potentiality}, covariant derivatives of \eqref{eq:HF:associator} yield further algebraic conditions, but these are algebraic consequences of \eqref{eq:HF:associator}, see \cite[sect.~7]{KSV2023}.

As a consequence, we are able to reconstruct the Hesse-Frobenius manifold, locally, from algebraic initial data: if we are provided with a commutative algebra with a Frobenius inner product and whose associator is consistent with \eqref{eq:HF:associator} at a point $q\in M$, then a local solution around that point exists, i.e.\ a local Hesse-Frobenius structure around $q$ that realises the initial algebra at $q$. This Hesse-Frobenius structure coincides with the original one.
In short, abundant systems arise, \emph{locally} for $n\geq3$, from initial data given by a commutative algebra with a compatible \innerproduct\ such that the associator condition~\eqref{eq:associator.cc} is respected at the initial point \cite{KSV2023,Vollmer2025_Frobenius,Vollmer_HF}.

\subsection{Superintegrable systems from Hesse-Frobenius structures}\label{sec:HF2SIS}

In this section, we consider Hesse-Frobenius manifolds $(M,g,\star)$ with an underlying metric $g$ of constant sectional curvature $\kappa$ that can be zero or non-zero. We explain how a second-order superintegrable system on $(M,g)$ can be obtained from this \emph{differential-geometric} data using the correspondence established in \cite{KSV2023,KSV2024,Vollmer2025_Frobenius,Vollmer_HF}:
these references establish that, on spaces of constant sectional curvature and for $n\geq3$, the data of a given Hesse-Frobenius structure on a simply connected manifold induces the data of an abundant system, and vice versa. The equations laid out in this section can be found in the cited references. To facilitate practical implementation, we formulate the approach in terms of a $3$-step procedure. The novelty here is that the correspondence between Hesse-Frobenius structures and superintegrable systems is later combined with additional techniques that allows us to obtain \emph{local} superintegrable systems \emph{explicitly} and \emph{systematically} directly from algebraic initial data.

Specifically, we will combine the $3$-step procedure presented in this section with an explicit Neumann series formula (see below, in Section~\ref{sec:sis.from.algebra}).
This allows us to construct local, \emph{flat} abundant systems from Frobenius algebras as initial data.
This, initially, will exclude the case with curvature. However, conification (Section~\ref{sec:conification}) will allow us to incorporate also the case with non-vanishing curvature.
\medskip

We denote the Hesse-Frobenius manifold by $(M,g,\star)$ and introduce the cubic $C$ as in \eqref{eq:defn.C}. We also introduce the shorthand $\hat C(X,Y)=\hat C(X)(Y)$ for brevity.
We use index notation to stress the local nature of the procedure.
It is shown in \cite{Vollmer2025_Frobenius,KSV2023} that, given a Hesse-Frobenius structure $(g,\star)$ of dimension $n\geq3$, a second-order maximally superintegrable system is obtained as follows:

\step{Step 1: Compute the structure tensor:}
The structure tensor is central in the description of abundant systems. We obtain it from the cubic $C_{ijk}$ of a Hesse-Frobenius structure by first computing the $(0,3)$-tensor field
\begin{equation}\label{eq:structure.tensor}
    T_{ijk}=3\left( C_{ijk}-\frac1n\,g_{ij}\,g^{ab}C_{abk}\right),
\end{equation}
and then the structure tensor
\begin{equation*}
    \hat T^k_{ij}= T_{ija}g^{ak}.
\end{equation*}
The abundant system can then be reconstructed from this structure tensor following step~2 and step~3.
Sections~\ref{sec:sis.from.algebra} and \ref{sec:direct.products} will address how to obtain $C_{ijk}$ from the initial data,  if $\kappa=0$. The case $\kappa\ne0$ is covered by Section \ref{sec:conification}.

\step{Step 2: Solve for the potential.}
With $\hat T_{ij}^k$ at hand, we integrate \eqref{eq:Wilczynski}, where $\nabla^2_{ij}V$ is the Hessian of $V$ with respect to the Levi-Civita connection of $g$, and where $\Delta$ denotes the Laplace-Beltrami operator.
By \cite{KSV2023}, the general solution of \eqref{eq:Wilczynski} depends on $n+2$ integration constants $c_j$, $1\leq j\leq n+2$.
Indeed, \eqref{eq:Wilczynski} leads to a first-order closed prolongation system of the form
\begin{subequations}\label{eq:Wilczynski.system}
\begin{align}
	\nabla^2_{ij}V&=\hat T_{ij}^a\,\partial_aV+\frac1n\,g_{ij}\,\Delta V
	\\
	\partial_k\Delta V &= \frac{n}{n-1}\,q_k^a\,\partial_aV+\frac{1}{n-1} t_k\,\Delta V
\end{align}
\end{subequations}
(see \cite{KSV2023}), where $t_k=\mathrm{trace}(\hat T)_k=\hat T_{ka}^a$ and $q_k^l=g^{ab}\,\partial_b\hat T_{ak}^l+g^{ab}\hat T_{ak}^c\hat T_{cb}^l-(n-1)\kappa\delta_k^l$.
The integrability conditions for \eqref{eq:Wilczynski.system} are obtained in \cite{KSV2023}.
In particular, a non-trivial computation \cite{KSV2023,Vollmer2025_Frobenius,Vollmer_HF} shows that the general analytic solution of \eqref{eq:Wilczynski.system} is determined by $n+2$ initial values if $C$ satisfies the conditions of a Hesse-Frobenius structure on $(M,g)$.

\step{Step 3: Compute the Killing tensors.}
The solution $V$ of \eqref{eq:Wilczynski.system} determines the superintegrable Hamiltonian \eqref{eq:Hamilton}. Due to Proposition~\ref{prop:functional.independence} below, based on a result in \cite{KSV2023}, it is indeed a second-order maximally superintegrable Hamiltonian. More precisely, for a generic choice of parameters, the potential $V$ is compatible -- in the sense of \eqref{eq:BD} -- with $2n-1$ \emph{functionally independent} integrals of the motion, including the Hamiltonian.
Thus, if we are only interested in the superintegrable Hamiltonian, we may stop here. However, in the light of Definition~\ref{defn:superintegrable.system}, one also desires to determine the associated integrals of the motion.
To this end, it suffices to provide the Killing tensor fields $K_{ij}$ associated to the system. Indeed, the corresponding integrals \eqref{eq:integral} are then obtained from \eqref{eq:preBD}, whose integrability condition is satisfied due to the properties of the Hesse-Frobenius structure \cite{KSV2023}. Specifically, the integrability condition for \eqref{eq:preBD} is the \emph{Bertrand-Darboux condition} \eqref{eq:bertrand-darboux}, which in the present context becomes, thanks to \eqref{eq:Wilczynski},
\begin{equation}\label{eq:BD}
    \left( \nabla_{[j}K^a_{i]}+K^b_{[i}\hat T_{j]b}^a\right)\partial_aV = 0\,,
\end{equation}
where square brackets denote skew symmetrisation in enclosed indices. We also recall that the $K_{ij}$ are the components of a Killing tensor field for $g$ such that \eqref{eq:integral} and~\eqref{eq:preBD} give a constant of the motion for \eqref{eq:Hamilton}.

We therefore now seek Killing tensor fields $K_{ij}$ such that \eqref{eq:BD} holds for any of the potentials $V$ obtained from solving \eqref{eq:Wilczynski.system} in the previous step.
As the metric $g$ is of constant sectional curvature, the solutions to the Killing equation~\eqref{eq:Killing} are known explicitly. Hence, they can be written as a linear combination of explicit Killing tensor fields with constant coefficients. In this way, \eqref{eq:BD} becomes a linear equation on these coefficients.

Summarising, any Killing tensor field for $g$ can be written in the form ($N=\frac{(n+1)!(n+2)!}{12(n-1)!n!}$, cf.\ \cite{Thompson1986})
\begin{equation*}
    K_{ij}(x) = \sum_{\nu=1}^N \beta_{\nu}K^{(\nu)}_{ij}(x)
\end{equation*}
for explicitly known Killing tensor fields $K^{(\nu)}_{ij}$.
In step~2, we found that the potential has the form ($M=n+2$)
\begin{equation*}
	V(x) = \sum_{\mu=1}^M a_{\mu}V^{(\mu)}(x)\,,
\end{equation*}
where $V^{(\mu)}$ are specific functions.
Hence, \eqref{eq:BD} becomes
\begin{equation}\label{eq:aux:BD.parametrised}
    \sum_{\mu,\nu} a_{\mu} \left( \beta_{\nu} g^{ab} K^{(\nu)}_{a[i,j]}+\beta_{\nu} g^{ak} K^{(\nu)}_{a[i}\hat T_{j]k}^b\right)\partial_bV^{(\mu)} = 0\,,
\end{equation}
where we denote covariant derivatives by a comma and use square brackets to denote skew-sym\-metri\-sation in the enclosed indices.
We seek the $\beta_\nu$ such that $K_{ij}$ is an associated Killing tensor of the system. Therefore, we need to require that \eqref{eq:aux:BD.parametrised} holds for any choice of the parameters $a_\mu$. We thus obtain the system
\begin{equation}\label{eq:shortcut.parametrised}
    \sum_{\nu} \beta_{\nu} \left( g^{ab} K^{(\nu)}_{a[i,j]}+g^{ak}K^{(\nu)}_{a[i}\hat T_{j]k}^b\right)\partial_bV^{(\mu)} = 0\,,\qquad \mu=1,\dots,n+2\,,
\end{equation}
which is a linear system of equations for $(\beta_1,\dots,\beta_N)$.

\begin{rmk}
	Since by assumption the integrability conditions of \eqref{eq:Wilczynski.system} hold generically, the Killing tensor fields associated to the system are equivalently obtained from the first-order closed prolongation system~\eqref{eq:shortcut}, cf.~\cite{KSV2023}. However, for the practical purposes here, it is often easier to solve the linear system~\eqref{eq:shortcut.parametrised} instead.
\end{rmk}

We mentioned earlier that, for a generic choice of solution to \eqref{eq:Wilczynski}, the integrals of the motion obtained through the system consisting of \eqref{eq:preBD} and \eqref{eq:shortcut} generically lead to a second-order maximally superintegrable system in the sense of Definition~\ref{defn:superintegrable.system}.
This is the statement of the next proposition.
\begin{prop}[\cite{KSV2023}]\label{prop:functional.independence}
	Let $(M,g,\star)$ be a simply connected Hesse-Frobenius manifold of dimension $n\geq2$. Let $C$ be the $(0,3)$-tensor field corresponding to $\star$ via $g$, and then let $\hat T$ be defined by \eqref{eq:structure.tensor}.
	Denote by $\mathcal K$ the space of solutions to \eqref{eq:shortcut}.
	For a \textit{generic} solution $V$ of \eqref{eq:Wilczynski.system}, one then finds $2n-2$ Killing tensor fields $K^{(k)}\in\mathcal K$ such that
	\[
		dH\wedge dF^{(1)}\dots\wedge dF^{(2n-2)}\ne0\,,
	\]
	where
	\[
		F^{(k)}(x,p)=(K^{(k)})_x^{ij}p_ip_j+W(x)
	\]
	with $W$ being a solution of \eqref{eq:preBD} for $K=K^{(k)}$.
	More precisely, the set of solutions to \eqref{eq:Wilczynski.system} that do not admit such functionally independent $(H,F^{(1)},\dots,F^{(2n-2)})$ is confined to an affine subspace with non-empty complement in the space of solutions to \eqref{eq:Wilczynski.system}.
\end{prop}
\begin{proof}
	The integrability conditions for \eqref{eq:Wilczynski.system}, \eqref{eq:shortcut}, and then for \eqref{eq:preBD}, are satisfied by the properties of a Hesse-Frobenius structure, compare \cite{KSV2023,Vollmer2025_Frobenius,Vollmer_HF}.
	Let $V$ be a solution of \eqref{eq:Wilczynski.system}, not necessarily a generic one.
	Furthermore, let $F^{(k)}$ denote $2n-2$ linearly independent integrals of the motion obtained by integrating \eqref{eq:shortcut} and \eqref{eq:preBD}, $1\leq k\leq 2n-2$.
	Now suppose that they are functionally dependent,
	\begin{equation}\label{eq:aux:functionally.dependent}
		dH\wedge dF^{(1)}\dots\wedge dF^{(2n-2)}=0.
	\end{equation}
	Then the proof of Lemma 6.2 in \cite{KSV2023} shows that
	\begin{equation}\label{eq:aux:affine.subspace}
		\partial_kV\in -\frac23\,{T_k}^{ab}p_ap_b+\ker\left(\sum_\ell\mu_\ell \hat K^{(\ell)}\right)\,,
	\end{equation}
	where $T_{ijk}=\hat T_{ij}^ag_{ak}$, $\hat K^{(\ell)}=g^{-1}K^{(\ell)}$, and where
	\[
		\mu_\ell=\frac{\partial\Phi}{\partial F^{(\ell)}}\circ F^{(\ell)}.
	\]
	Indeed, substitute \eqref{eq:integral} and then \eqref{eq:preBD} into \eqref{eq:aux:functionally.dependent}. A short computation yields
	\[
		\sum_\ell \mu_\ell K^{(\ell)\,k}_{j} \left(
			\frac23 {T_k}^{ab}p_ap_b +\partial_kV
		\right) = 0\,,
	\]
	invoking \eqref{eq:shortcut}. The claim follows.
\end{proof}

Because of Proposition~\ref{prop:functional.independence}, and by a slight abuse of notation, we call a linearly parametrised family of solutions to~\eqref{eq:Wilczynski} a potential, as well.

\begin{defn}
	For a Hesse-Frobenius manifold $(M,g,\star)$, we call the abundant system defined via \eqref{eq:structure.tensor}, \eqref{eq:Wilczynski.system} and~\eqref{eq:shortcut} its \emph{corresponding superintegrable system}.
\end{defn}

For simplicity, if the specific choice of the integrals $F^{(k)}$ in Definition~\ref{defn:superintegrable.system} is not important, we are sometimes going to specify only the superintegrable Hamiltonian. If the underlying metric is clear, we may even just specify the superintegrable potential, as there will be no risk of confusion.

\subsection{Superintegrable systems from Frobenius algebras}\label{sec:sis.from.algebra}

In this section we assume $\kappa=0$.
Note that the procedure outlined in Section~\ref{sec:HF2SIS} commences with a differential-geometric structure, namely a Hesse-Frobenius structure.
However, \eqref{eq:potentiality.flat} is a first-order closed non-linear prolongation system for $C$. Hence, the algebraic information $C_q$ in a point $q\in M$ suffices to reconstruct $C$ locally, since the $C_{ijk}$ are analytic around $q$.
In this section, we carry out this local construction.
Starting from a Frobenius algebra as initial data at a point $q\in M$,
we seek the flat Hesse-Frobenius structure locally around $q$.
The $3$-step procedure in Section~\ref{sec:HF2SIS} then allows us to recover
the corresponding flat second-order superintegrable systems locally.
\medskip

Consider $\mathbb K^n$ with the \innerproduct\ $\rho$ and standard addition $+$, as well as a commutative and associative product $\blackdiamond$ such that the Frobenius property $\rho(u\blackdiamond v,w)=\rho(u,v\blackdiamond w)$ is satisfied ($u,v,w\in\mathbb K^n$). We introduce $c(u,v,w)=\rho(u\blackdiamond v,w)$, which is totally symmetric.
The Frobenius algebra now serves as the initial data for the system of partial differential equations~\eqref{eq:potentiality.flat}, subject to \eqref{eq:WDVV},
for the totally symmetric cubic tensor field $C_{ijk}$ on the manifold $M$, locally around a point $q\in M$.

We are going to use a Neumann expansion to obtain the unique local analytic solution
around~$q$ for the cubic $C$ with the prescribed initial algebraic structure.
To this end, we first introduce the endomorphisms $c_j$ on $T_qM\simeq\mathbb K^n$ given by
\begin{equation*}
	\rho(c_j(v),w)=c(e_j,v,w)
\end{equation*}
for $j\in\{1,2,\dots,n\}$ with respect to a basis $(e_1,\dots,e_n)$ of $\mathbb K^n$.
Now consider local affine coordinates $(x^1,\dots,x^n)$ around $q\in M$, centered at $q$, defining $c_j$ as above in the obvious manner.
For $j\in\{1,2,\dots,n\}$, we then define
\begin{equation}
	R(x)=\left(\mathrm{id}-\sum_{j=1}^n x^ic_i\right)^{-1}
\end{equation}
on the neighborhood of the origin where $\det(\mathrm{id}-\sum_{j=1}^n x^ic_i)\ne0$ within the domain admitted by the local coordinates $(x^j)$.
In the next lemma, we treat the $x^j$ as parameters.

\begin{lem}\label{lem:R(x)}
	$R(x)$ is self-adjoint with respect to $\rho$, and it belongs to the centroid of the algebra $(\mathbb K^n,\rho,\blackdiamond)$.
\end{lem}
\begin{proof}
	From the Frobenius property of $\blackdiamond$ with respect to $\rho$ it follows that the $c_j$ are self-adjoint with respect to $\rho$. Since $\blackdiamond$ is associative, the $c_j$ mutually commute and hence we infer that also $\mathrm{id}-\sum_{j=1}^n x^ic_i$ is self-adjoint with respect to $\rho$.
	It follows that $R(x)$ is self-adjoint with respect to $\rho$.
	
	The centroid is the algebra of linear maps $\tau$ on $\mathbb K^n$ that satisfy $\tau(v\blackdiamond w)=\tau(v)\blackdiamond w=v\blackdiamond\tau(w)$.
	Now, since
	$c_j(v\blackdiamond w)
		= (c_jv)\blackdiamond w
		= v\blackdiamond(c_jw)$
	due to the commutativity and associativity of~$\blackdiamond$, it is clear that each $c_j$ belongs to the centroid, and hence also $R(x)$ and $R(x)^{-1}$.
\end{proof}

We now employ $R(x)$ to extend the initial Frobenius algebra $(T_qM\simeq\mathbb K^n,\rho,\blackdiamond)$ to a neighborhood of $q$, defining, in local coordinates,
\[
	\hat C_x(v,w)=R(x)(v\blackdiamond w),
\]
where we identify $TM$ in a maybe shrunk neighborhood around $q$ with the trivial bundle $\mathbb K^{2n}\to\mathbb K^n$.
\begin{lem}\label{lem:hatC_x}
	$\hat C$ defines a Hesse-Frobenius structure, i.e.\ a commutative and associative product that has the Frobenius property and satisfies \eqref{eq:potentiality.flat}.
\end{lem}
\begin{proof}
	The commutativity of $\hat C$ follows immediately from that of $\blackdiamond$, and the associativity of $\hat C$ is obtained from the associativity of $\blackdiamond$ using also the fact that $R(x)$ belongs to the centroid of $(\mathbb K^n,\rho,\blackdiamond)$. For the Frobenius property, we similarly compute
	\[
		\rho(u,R(x)(v\blackdiamond w))
		= \rho(R(x)(u),v\blackdiamond w)
		= \rho(R(x)(u)\blackdiamond v,w)
		= \rho(R(x)(u\blackdiamond v),w),
	\]
	using Lemma~\ref{lem:R(x)} and the Frobenius property of $\blackdiamond$ with respect to $\rho$.
	Finally, we obtain
	\[
		(\nabla_k\hat C)_x(v,w)=(\partial_{x^k}R(x))(v,w)
		=R(x)c_kR(x)(v,w)=\hat C(e_k,\hat C(v,w))
	\]
	since $0=\partial_{x^k}(R(x)^{-1}R(x))=-c_kR(x)+R(x)^{-1}\partial_{x^k}R(x)$.
\end{proof}

As a result of Lemmas~\ref{lem:R(x)} and~\ref{lem:hatC_x}, we obtain the unique analytic solution of~\eqref{eq:potentiality.flat} defining a local Frobenius manifold structure around $q$ with the initial Frobenius algebra $(T_qM\simeq\mathbb K^n,\rho,\blackdiamond)$: locally around $q$ we have $g$ given by $g(\partial_i,\partial_j)=\rho(e_i,e_j)$ using the identification $T_qM\simeq\mathbb K^n$ and local coordinates $x=(x^1,\dots,x^n)$ around $q$. For each component of $C_{ij}^\ell$ (with $i,j,\ell\in\{1,2,\dots,n\}$), we then have the Neumann expansion
\begin{equation}\label{eq:ansatz}
	C_{ij}^\ell = \sum_{\mu=0}^\infty \sum_{|\alpha|=\mu} \frac{\mu!}{\alpha!} x^\alpha\,\,(c_1^{\alpha_1}\cdots c_n^{\alpha_n}\sigma_{ij})^\ell
\end{equation}
where $\alpha=(\alpha_1,\dots,\alpha_n)$ is a multi-index, such that $x^\alpha=\prod_{j=1}^n (x^j)^{\alpha_j}$.
We have silently introduced $\sigma_{ij}$ as the vector valued map $\mathbb K^n\times\mathbb K^n\to\mathbb K^n$, $(u,v)\mapsto c_{ij}^ku^iv^j\,e_k$, encoding $\blackdiamond$.
Finally, we introduce $C_{ij}^k=g^{ka}C_{ija}$.
Note that, by construction, the derivative of $C_{ij}^\ell$ at $q$ is
\begin{equation*}
	(\nabla_kC_{ij}^\ell)_q = C_{ka}^\ell(q)C_{ij}^a(q) = c_{ka}^\ell c_{ij}^a = (c_k\sigma_{ij})^\ell,
\end{equation*}
consistent with \eqref{eq:potentiality.flat}. Higher derivatives of $C^\ell_{ij}$ at $q$ are then obtained iteratively, repeatedly using~\eqref{eq:potentiality.flat}.
We conclude:
\begin{prop}
	Let $(M,g)$ be a flat manifold, $q\in M$, and $(\mathbb K^n\simeq T_qM,\rho=g_q,\blackdiamond)$ a Frobenius algebra. Then there exists a small neighborhood $U$ around $q$ with a unique Hesse-Frobenius structure $(U,g|_U,\star)$ consistent with the initial data $(\mathbb K^n\simeq T_qM,\rho,\blackdiamond)$ at $q$. This unique analytic solution is given by Formula~\eqref{eq:ansatz}, i.e.\ by $\hat C$ in Lemma~\ref{lem:hatC_x}.
\end{prop}

\subsection{Direct products}\label{sec:direct.products}

In this section, we still assume $\kappa=0$.
We now describe a rather basic construction for direct products of flat Hesse-Frobenius manifolds. However, this construction leads to a non-obvious construction on the side of superintegrable systems. In particular, the superintegrable system corresponding to the product Hesse-Frobenius manifold admits non-obvious additional associated Killing tensor fields.
Therefore, as we are going to see later, in spite of its simplicity on the side of Hesse-Frobenius structures, this construction is quite useful for understanding the corresponding superintegrable systems. It also provides an efficient method to construct abundant systems.
\medskip

We assume that $(U,g,\star)$ and $(W,h,\ast)$ are two \emph{flat} Hesse-Frobenius manifolds.
We then consider the direct product $M=U\times W$ with the metric $G=g+h$ and together with the product $\divideontimes=\star+\ast$.
Here, we understand $g$ and $h$ as well as $\star$ and $\ast$ to be extended trivially to $M$, such that
\[
	G=\begin{pmatrix} g & 0 \\ 0 & h \end{pmatrix}\qquad\text{and}\qquad \star\circ\,\ast=\ast\circ\star=0
\]
It is easily checked that $\divideontimes$ is commutative and associative, and that it has the Frobenius property with respect to $G$. Moreover, we find that
\[
	\nabla^G(\star+\ast)=\nabla^g\star+\nabla^h\ast
	= \star\circ\star+\ast\circ\ast=(\star+\ast)\circ(\star+\ast)\,,
\]
where $\nabla^G$, $\nabla^g$ and $\nabla^h$ denote the Levi-Civita connections of $G,g$ and $h$, respectively.
We furthermore obtain that
$ (M,G,\divideontimes)$
is also a Hesse-Frobenius structure. It follows, see Section~\ref{sec:HF2SIS}, that it gives rise to a second-order superintegrable system on $(M,G)$.

\begin{prop}
	Assume that $(U,g,\star)$ and $(W,h,\ast)$ are flat Hesse-Frobenius manifolds with
	dimension $n_U$ and $n_W$, respectively.
	Via step~2 of Section~\ref{sec:HF2SIS}, they carry abundant systems whose superintegrable potentials we denote by $V_U(x)$ and $V_W(y)$, respectively, where $(x,y)$ are adapted coordinates on $M=U\times W$.
	Then there exists a $(n_U+n_W+2)$-parameter potential
	\begin{equation}\label{eq:glued.potential}
		V_M(x,y)=\widetilde{V}_U(x)+\widetilde{V}_W(y)\,,
	\end{equation}
	such that the Hamiltonian $G^{ij}(x,y)p_ip_j+V_M(x,y)$ on $T^*M$ carries an abundant system.
\end{prop}

We comment on the number of parameters of the potentials $V_U$, $V_W$ and $V_M$: the abundant systems corresponding to $(U,g,\star)$ and $(W,h,\ast)$ have potentials involving $N_U=n_U+2$ and $N_W=n_V+2$ parameters, respectively. Hence, the naive sum $V_U(x)+V_W(y)$ contains $n_U+n_V+4$ parameters. This is more than the number of $n_U+n_W+2$ parameters, which we expect for the potential $V_M(x,y)$ of the abundant system corresponding to the direct product Hesse-Frobenius manifold $(M,G,\divideontimes)$.
The proof identifies the additional constraints on the parameters.

\begin{proof}
	We obtain $\widetilde{V}_U$ and $\widetilde{V}_W$ from $V_U$ and $V_W$, respectively, imposing restrictions on the parameters.	
	Consider the sum $V_U(x)+V_W(y)$ with, a priori, $n_U+n_W+4$ free parameters.
	We first observe that in each set of the $N_U$ and $N_W$ parameters in $V_U$ and $V_W$, respectively, one parameter is only an additive constant. We conclude that $V_M$ can have, at most, $n_U+n_W+3$ independent parameters. However, the potential $V_M$ is subject to the condition~\eqref{eq:Wilczynski}, which imposes one additional restriction on the parameters. More specifically, we have the condition
	\begin{equation}\label{eq:constraint.on.glued.potential}
		\nabla^{G\,2}_{ij}V_M=\hat T_{ij}^k\,\partial_kV_M+\frac1n\,G_{ij}\,\Delta_G V_M\,,
	\end{equation}
	where $\Delta_G$ is the Laplace-Beltrami and $\nabla^{G\,2}$ the Hessian operator for $G$. Similarly, we denote the Laplace-Beltrami operators for $g$ and $h$, respectively, by $\Delta_g$ and $\Delta_h$. The respective Hessians are $\nabla^{g\,2}$ and $\nabla^{h\,2}$.
	Denote the cubic tensor field for~$\star$ by~$C$ and that for~$\ast$ by~$B$. Then \eqref{eq:constraint.on.glued.potential} reads, in terms of $C_{ij}^k $ and $B_{ij}^k$,
	\begin{multline*}
		\nabla_{ij}^{g\,2}V_U+\nabla_{ij}^{h\,2}V_W
		=
		3C_{ij}^{k}\partial_kV_U
		+ 3B_{ij}^{k}\partial_kV_W
		\\
		- \frac{3}{n_U}g_{ij}c^{k}\partial_kV_U
		- \frac{3}{n_W}h_{ij}b^{k}\partial_kV_W
		+\frac{1}{n} (g_{ij}+h_{ij}) (\Delta_gV_U+\Delta_hV_W)\,,
	\end{multline*}
	where we trivially extend $g,h,C$ and $B$ to $M$ and introduce $c_k=g^{ab}C_{abk}$ and $b_k=h^{ab}B_{abk}$.
	One then easily verifies that \eqref{eq:constraint.on.glued.potential} is equivalent to
	\begin{subequations}\label{eqs:glued.Wilczynski.decomposed}
		\begin{align}
			\label{eq:glued.Wilczynski.decomposed.U}
			\nabla_{ij}^{g\,2}V_U
			&= 3C_{ij}^{k}\partial_kV_U
			- \frac{3}{n_U}g_{ij}c^{k}\partial_kV_U
			+\frac{1}{n} g_{ij} (\Delta_gV_U+\Delta_hV_W),
			\\
			\label{eq:glued.Wilczynski.decomposed.W}
			\nabla_{ij}^{h\,2}V_W
			&=
			3B_{ij}^{k}\partial_kV_W
			- \frac{3}{n_W}h_{ij}b^{k}\partial_kV_W
			+\frac{1}{n} h_{ij} (\Delta_gV_U+\Delta_hV_W).
		\end{align}
	\end{subequations}
	It is also straightforwardly confirmed that the trace-free parts of \eqref{eq:glued.Wilczynski.decomposed.U} and \eqref{eq:glued.Wilczynski.decomposed.W} are implied by the conditions \eqref{eq:Wilczynski}, evaluated for the respective second-order superintegrable systems on $U$ and $V$. We are therefore left with the traces
	\begin{subequations}\label{eqs:glued.Wilczynski.decomposed.traces}
		\begin{align}
			\label{eq:glued.Wilczynski.decomposed.trace.U}
			\Delta_gV_U
			&
			= \frac{n_U}{n} (\Delta_gV_U+\Delta_hV_W),
			\\
			\label{eq:glued.Wilczynski.decomposed.trace.W}
			\Delta_hV_W
			&
			= \frac{n_W}{n} (\Delta_gV_U+\Delta_hV_W).
		\end{align}
	\end{subequations}
	We conclude that \eqref{eqs:glued.Wilczynski.decomposed.traces} impose one additional restriction on the parameters in $V_U(x)+V_W(y)$, leading to $\widetilde{V}_U(x)+\widetilde{V}_W(y)$ in \eqref{eq:glued.potential}.
	Indeed, observe that the conditions \eqref{eqs:glued.Wilczynski.decomposed.traces} are not independent, which is immediately verified by forming their sum.
	Therefore, we are left with one independent constraint on the $n_U+n_W+3$ parameters of $V_U(x)+V_W(y)$, leading precisely to the expected number of $n_U+n_W+2$ parameters in $V_M(x,y)$.
\end{proof}

\subsection{Conification}\label{sec:conification}

We now assume $\kappa\ne0$.
In this section, we explain how an abundant system on a space of non-zero constant sectional curvature $\kappa$ can be lifted to an abundant system corresponding to a flat Hesse-Frobenius structure on the flat cone over the original system. Thereby, the techniques in Sections~\ref{sec:sis.from.algebra} and~\ref{sec:direct.products} become available also for Hesse-Frobenius structures with curvature.

As a disclaimer, we remark that the conification construction of the following proposition is not new. In fact, it turns out to be a special case of the warped products discussed recently in chapter~8 of~\cite{BKM2023}. This, however, is a feature rather than a drawback: indeed, it shows the power of the general framework inherited here from \cite{KSV2023,Vollmer2025_Frobenius,Vollmer_HF}. Note that \cite{BKM2023} does not discuss superintegrability, but Nijenhuis geometry, Frobenius pencils and Poisson structures. Thanks to the correspondence of abundant systems and Hesse-Frobenius structures, we can transfer it to the realm of superintegrability.

In what follows, we denote the Lie derivative by $\mathfrak L$.
\begin{prop}\label{prop:conification}
	Let $(N,h,\ast)$ be a Hesse-Frobenius manifold of constant sectional curvature $\kappa\ne0$.
	Let $M=(0,1)\times N$ be the flat cone over $N$ with the cone metric $g = dr^2+\kappa\,r^2\,h$.
	Then
	\begin{subequations}\label{eqs:product.conditions}
		\begin{align}
				X\star Y &= X\ast Y-r\,\kappa\,h(X,Y)\,\partial_r
				\label{eq:product.conditions.horizontal} \\
				X\star\partial_r &= -\frac1r\,X
				\label{eq:product.conditions.mixed} \\
				\partial_r\star\partial_r &= -\frac1r\,\partial_r
				\label{eq:product.conditions.vertical}
			\end{align}
	\end{subequations}
	defines a flat unital Hesse-Frobenius structure $\star$ on $(M,g)$ with the unit $\mathsf{e}=-r\partial_r$ that satisfies $\mathfrak L_{\mathsf{e}}g=-2g$.
\end{prop}
For later reference, we introduce the following terminology.
\begin{defn}\label{defn:conification}
	Let $(N,h,\ast)$ be a Hesse-Frobenius manifold of constant sectional curvature $\kappa\ne0$. Then we call $(M,g,\star,\mathsf{e})$ its \emph{flat Hesse-Frobenius conification}, where $M=(0,1)\times N$ and $g=dr^2+\kappa r^2h$, $\mathsf{e}=-r\partial_r$ and where $\star$ is defined by~\eqref{eqs:product.conditions}, as in Proposition~\ref{prop:conification}.
	
	Moreover, let $\mathcal S_N$ denote the superintegrable system corresponding to $(N,h,\ast)$, and $\mathcal S_M$ the superintegrable system corresponding to $(M,g,\star)$. We then call $\mathcal S_M$ the \emph{flat conification} of $\mathcal S_N$.
\end{defn}
Note that it suffices to formulate the statement of Proposition~\ref{prop:conification} on the side of Hesse-Frobenius structures. We prove it by an explicit construction. As mentioned, a proof can also be found in \cite{BKM2023}, albeit in a different language.
\begin{proof}[Proof of Proposition~\ref{prop:conification}]
	By the hypothesis, $(N,h,\ast)$ is a $n$-dimensional Hesse-Frobenius structure of constant sectional curvature $\kappa\ne0$.
	We recall that the Levi-Civita connection $\nabla^M$ of $g$ is given, for $X,Y\in\mathfrak X(N)\subset\mathfrak X(M)$, by
	\begin{equation}\label{eq:nabla.M.horizontal}
		\nabla^M_XY = \nabla^N_XY-\kappa\,r\,h(X,Y)\partial_r\,,
	\end{equation}
	where $\nabla^N$ is the Levi-Civita connection of $h$.
	For $X\in\mathfrak X(N)$ and $u\in\mathfrak X(U)$,
	\begin{equation}\label{eq:nabla.M.mixed}
		\nabla^M_uX=\nabla^M_Xu = \frac1r\,u(dr)\,X.
	\end{equation}
	Finally, for $u,v\in\mathfrak X(U)$,
	\begin{equation}\label{eq:nabla.M.vertical}
		\nabla^M_uv = \nabla^U_uv = u^k\partial_k(u^i)\,\partial_i\,,
	\end{equation}
	where $\nabla^U_uv$ denotes the directional derivative of $v$ with respect to $u$, i.e.\ the Levi-Civita connection of $dr^2$.
	Our objective is to lift the Hesse-Frobenius structure on $N$ to the flat cone $M$ over $N$. To this end, let us introduce, on $N$, the symmetric $(0,3)$-tensor field $\Psi$ and its associated $(1,2)$-tensor field $\hat\Psi$ by
	\[
		\Psi(X,Y,Z) = g(\hat\Psi(X,Y),Z) = g(X\ast Y,Z).
	\]
	Note that $\hat\Psi$ is just an auxiliary notation for $\ast$ that will be handy in the following computations.
	A direct computation, using \eqref{eq:nabla.M.horizontal}, yields
	\begin{align}
		[\hat\Psi(X),\hat\Psi(Y)]Z &= -R^N(X,Y)Z = \kappa\,(h(X,Z)Y-h(Y,Z)X)
		\label{eq:associator.cc}
		\\
		\nabla^N_Z\hat\Psi(X,Y) &= \hat\Psi(Z,\hat\Psi(X,Y))+\kappa\,(2\,h(X,Y)Z+h(X,Z)Y+h(Y,Z)X)
		\label{eq:potentiality.cc}
	\end{align}
	where $R^N$ is the curvature tensor for $\nabla^N$.
	We now introduce a symmetric $(0,3)$-tensor field $C$ on~$M$ by decreeing the ansatz
	\begin{align*}
		C(X,Y,Z)&=\kappa r^2 \Psi(X,Y,Z)\,,
		&
		C(X,Y,\partial_r)&=\alpha r^\nu h(X,Y)\,,
		\\
		C(X,\partial_r,\partial_r)&=0\,,
		&
		C(\partial_r,\partial_r,\partial_r)&=\beta r^\mu\,,
	\end{align*}
	where $X,Y,Z$ are horizontal vector fields (on $N$), and where $\nu,\mu,\alpha,\beta$ are constants to be determined in such a manner that $C$ defines a flat Hesse-Frobenius structure on $(M,g)$. Let $\hat C=g^{-1}C$ be the $(1,2)$-tensor field corresponding to $C$, thanks to $g$.
	
	In order to determine the parameters $\nu,\mu,\alpha,\beta$ in our ansatz, we choose coordinates\footnote{From now on, we label the coordinates $x_0,\dots,x_n$ by subscripts, rather than superscripts, to ensure a concise notation and to prevent confusion between coordinate names and exponents.}%
	$(x_1,\dots,x_n)$ on $N$, which in index notation we represent by small Latin letters $a,b$ etc. On $M$, we then use coordinates $(x_0=r,x_1,\dots,x_n)$, and we represent them by capital Latin letters. The index $0$ refers to $x_0$.
	For the components of the associator of the product $\hat C$, we find:
	\begin{align*}
		\mathrm{Assoc}_{IJKL}
		&= g^{AB}\left( C_{IJA}C_{KLB}-C_{IKA}C_{LJB}\right)
		\\
		&= \left( C_{IJ0}C_{KL0}-C_{IK0}C_{LJ0}\right)
		+\frac{h^{ab}}{\kappa r^2}\,\left( C_{IJa}C_{KLb}-C_{IKa}C_{JLb} \right).
	\end{align*}
	It follows that
	\begin{align*}
		\mathrm{Assoc}_{ijkl}
		&= \left( C_{ij0}C_{kl0}-C_{ik0}C_{jl0} \right)
		+g^{ab}\left( C_{ija}C_{klb}-C_{ika}C_{jlb}\right)
		= (\alpha^2 r^{2\nu} - \kappa^2 r^2)\left( h_{ij}h_{kl}-h_{ik}h_{jl} \right),
		\\
		\mathrm{Assoc}_{0jkl}
		&= \left( C_{j00}C_{kl0}-C_{k00}C_{jl0} \right)
		+g^{ab}\left( C_{ja0}C_{klb}-C_{ka0}C_{jlb}\right)
		= \alpha r^\nu\,h^{ab} \left( h_{ja}\Psi_{klb}-h_{ka}\Psi_{jlb} \right)=0,
		\\
		\mathrm{Assoc}_{00kl}
		&= \left( C_{000}C_{kl0}-C_{ik0}C_{l00}\right)
		+g^{ab}\left( C_{a00}C_{klb}-C_{ka0}C_{lb0}\right)
		= \alpha r^\nu\left( \beta r^\mu-\frac{\alpha}{\kappa} r^{\nu-2} \right) h_{kl}\,.
	\end{align*}
	We conclude that $\nu=1,\mu=-1,\alpha=\varepsilon\kappa,\beta=\varepsilon,\varepsilon\in\{\pm1\}$ ensures $\mathrm{Assoc}=0$, this is it guarantees that \eqref{eq:associator.cc} implies the associativity of the product given by $\hat C$.
	
	It remains to ensure that \eqref{eq:potentiality.flat} is satisfied, provided \eqref{eq:potentiality.cc} holds. We claim that this is the case for the choice $\varepsilon=-1$, i.e.\ for $\alpha=-\kappa,\beta=-1$ (one easily disproves the other solution branch).	
	Our ansatz then specifies to
	\begin{align*}
		C(X,Y,Z)&=\kappa r^2 \Psi(X,Y,Z)\,,
		&
		C(X,Y,\partial_r)&=-\kappa r h(X,Y)\,,
		\\
		C(X,\partial_r,\partial_r)&=0\,,
		&
		C(\partial_r,\partial_r,\partial_r)&=-\frac1r\,.
	\end{align*}
	To verify our claim, we compute the covariant derivatives of $C$, using our ansatz. We find
	\begin{align*}
		\nabla^M_0C_{000}&= \frac{1}{r^2}\,,
		&
		\nabla^M_0C_{00i}&= 0\,,
		&
		\nabla^M_0C_{0ij}&= \kappa\,h_{ij}\,,
		&
		\nabla^M_0C_{ijk}&= -\kappa r \Psi_{ijk}\,,
		\\
		\nabla^M_lC_{0ij}&= -\kappa r \Psi_{ijl}\,,
		&
		\nabla^M_lC_{00j}&= 0\,,
		&
		\nabla^M_lC_{000}&= 0
	\end{align*}
	and
	\[
	\nabla^M_lC_{ijk} = \kappa r^2 \nabla^N_l\Psi_{ijk}
	-\kappa^2 r^2 (h_{jk}h_{il}+h_{ik}h_{jl}+h_{ij}h_{kl})\,.
	\]
	(We observe that $C$ satisfies the potentiality property, i.e.\ that $\nabla^MC$ is totally symmetric.)
	On the other hand,
	\begin{align*}
		g^{AB}C_{00A}C_{00B} &= \frac{1}{r^2}\,,
		&
		g^{AB}C_{00A}C_{0iB} &= 0\,,
		\\
		g^{AB}C_{0Ai}C_{0Bj} &= \kappa h_{ij}\,,
		&
		g^{AB}C_{0kA}C_{ijB} &= -\kappa r \Psi_{ijk}\,,
		\\
		g^{AB}C_{0lA}C_{ijB} &= -\kappa r \Psi_{ijl}\,,
		&
		g^{AB}C_{0lA}C_{0jB} &= \kappa h_{lj}\,,
		\\
		g^{AB}C_{00A}C_{0lB} &= 0\,,
		&
		g^{AB} C_{00A} C_{ljB} &= \kappa^2 h_{jl}\,,
	\end{align*}
	as well as
	\[
	g^{AB}C_{ilA}C_{jkB} = \kappa^2 r^2 h_{il} h_{jk}+\kappa r^2 \Psi_{ila}h^{ab}\Psi_{jkb}\,.
	\]
	We conclude that \eqref{eq:potentiality.flat} holds for our ansatz.
	In summary, we have obtained that the product defined by
	\begin{align}
		\hat C(X,Y)&=\hat\Psi(X,Y)-\kappa r h(X,Y)\partial_r\,,
		&
		\hat C(X,\partial_r)&=-\frac1r\,X\,,
		&
		\hat C(\partial_r,\partial_r)&=-\frac1r\,\partial_r
	\end{align}
	($X,Y$ are horizontal vector fields) defines a flat Hesse-Frobenius structure on the cone over $N$, i.e.\ on~$(M,g)$.
	In our coordinates,
	\begin{align*}
		C_{ij}^{k} &= \Psi_{ij}^{k}\,,
		&
		C_{ij}^{0} &= -\kappa r h_{ij}\,,
		&
		C_{i0}^{k} &= -\frac1r\,\delta_i^k\,,
		&
		C_{i0}^{0} &= 0\,,
		&
		C_{00}^{k} &= 0\,,
		&
		C_{00}^{0} &= -\frac1r\,.
	\end{align*}
	Note that $\mathsf{e}=-r\partial_r$ is a unit, i.e.\ $\star$ is unital. Indeed, $\mathsf{e}\star = \mathrm{id}_{\mathfrak X(M)}$.
	Moreover, the Lie derivative of $g$ with respect to $\mathsf{e}$ is $\mathfrak L_{\mathsf{e}}g=-2g$.
\end{proof}

We also give a converse statement.

\begin{prop}\label{prop:reverse.conification}
	Let $(M,h)$ be a manifold of constant sectional curvature $\kappa\ne0$, and let $M=(0,1)\times N$ be the cone over $N$ equipped with the flat metric $g=dr^2+\kappa r^2 h$.
	Assume further that $(M,g)$ be equipped with a flat unital Hesse-Frobenius structure $\star$ with unit $\mathsf{e}=-r\partial_r$, such that the $(1,2)$-tensor field $\hat\Psi$, defined by
	\begin{equation}\label{eq:hatPsi.via.hatC}
		\hat\Psi(X,Y)=X\star Y-\kappa\,h(X,Y)\,\mathsf{e}
	\end{equation}
	(for horizontal vector fields $X,Y$) is horizontal and independent of $r$.
	Then $(N,h,\hat\Psi)$ defines a Hesse-Frobenius structure on the base.
\end{prop}
This proposition reduces the classification problem for Hesse-Frobenius structures to the flat case. Moreover, it allows us, in Section~\ref{sec:SW.and.sphere}, to relate a known $n$-dimensional abundant system with positive constant sectional curvature to an abundant system on $(n+1)$-dimensional flat space.\bigskip

\begin{proof}[Proof of Proposition~\ref{prop:reverse.conification}]
	We consider the unital product $\star$ on the cone $M$ over~$N$ with metric~$g$.
	Note that $\mathsf{e}=-r\partial_r$ implies that $\mathfrak L_{\mathsf{e}}g=-2g$ and $g(\mathsf{e},\mathsf{e})=r^2$, where $r\in(0,1)$.
	By the hypothesis, we furthermore have that $\mathsf{e}\star = \mathrm{id}_{\mathfrak X(M)}$ and that, for horizontal vector fields $X,Y$,
	\[
		g(\hat\Psi(X,Y),\mathsf{e})=0.
	\]
	Denote the Levi-Civita connections of $g$ and $h$ by $\nabla^M$ and $\nabla^N$, respectively. We recall the formulas \eqref{eq:nabla.M.horizontal}, \eqref{eq:nabla.M.mixed} and~\eqref{eq:nabla.M.vertical} for the relationship of these connections.
	We are also going to keep our previous notation of referring to local coordinates on $N$ by small Latin letters, and using capital Latin letters for the corresponding local coordinates on $M$, where $0$ stands for $\partial_r$ or $dr$, respectively.
	
	Our objective is to prove that $\hat\Psi$ defines a commutative product $\ast$ on $\mathfrak X(N)$ such that the Frobenius property holds, the potentiality condition~\eqref{eq:potentiality.cc} is fulfilled and the associator satisfies
	\begin{equation}
		\mathrm{assoc}(X,Y,Z) = (X\ast Y)\ast Z - X\ast(Y\ast Z)
		= \kappa \left( h(Y,Z)X - h(X,Y)Z \right).
	\end{equation}
	To this end, let $X,Y,Z$ be horizontal vector fields.
	We first verify that the $(0,3)$-tensor field $\Psi=h\hat\Psi$ is totally symmetric.
	Indeed,
	\[
		\Psi(X,Y,Z) = h(\hat\Psi(X,Y),Z)
		= \frac{1}{\kappa r^2}\left( g(X\star Y,Z)-g(\mathsf{e},Z)h(X,Y) \right)
		= \frac{1}{\kappa r^2} C(X,Y,Z)
	\]
	is symmetric in all three arguments.
	Next, we compute the associator of $\ast$,
	\begin{align*}
		\mathrm{assoc}(X,Y,Z)
		&= \hat\Psi(\hat\Psi(X,Y),Z)-\hat\Psi(X,\hat\Psi(Y,Z))
		\\
		&= \hat\Psi(X,Y)\star Z -\kappa\mathsf{e} h(\hat\Psi(X,Y),Z)
			-X\star \hat\Psi(Y,Z) +\kappa\mathsf{e} h(X,\hat\Psi(Y,Z))
		\\
		&= \left( X\star Y-\kappa\mathsf{e}h(X,Y) \right)\star Z -\kappa\mathsf{e} h(\hat\Psi(X,Y),Z)
		\\ &\qquad
		-X\star\left( Y\star Z-\kappa\mathsf{e}h(Y,Z) \right) +\kappa\mathsf{e} h(X,\hat\Psi(Y,Z))
		\\
		&= \kappa \left( h(Y,Z)X-h(X,Y)Z \right) +\mathsf{e} \left(h(X,\hat\Psi(Y,Z)) -h(\hat\Psi(X,Y),Z)\right)
		\\
		&= \kappa \left( h(Y,Z)X-h(X,Y)Z \right)\,.
	\end{align*}
	Finally, we verify \eqref{eq:potentiality.cc}: let $X,Y,Z,W$ be horizontal vector fields and recall that $\hat C(X,Y)=X\star Y$.
	We compute, using \eqref{eq:nabla.M.horizontal} and \eqref{eq:hatPsi.via.hatC},
	\begin{align*}
		\kappa r^2 h\left( (\nabla^N_Z\hat\Psi)(X,Y),W \right)
		&= g\left(
			\nabla^N_Z(\hat\Psi(X,Y))-\hat\Psi(\nabla^N_ZX,Y)-\hat\Psi(X,\nabla^N_ZY),
			W \right)
		\\
		&= g\left(\nabla^M_Z(\hat\Psi(X,Y)),W \right)
		\\ &\qquad
		-g\left(\hat C(\nabla^N_ZX,Y),W \right)
		-g\left(\hat C(X,\nabla^N_ZY),W \right)
		\\
		&= g\left( (\nabla^M_Z\hat C)(X,Y),W \right)
		-\kappa h(X,Y) g\left(\nabla^M_Z\mathsf{e},W \right)
		\\ &\qquad
		+\kappa h(Z,X)\,g\left(\hat C(\mathsf{e},Y),W \right)
		+\kappa h(Z,Y)\,g\left(\hat C(X,\mathsf{e}),W \right)\,.
	\end{align*}
	Because of \eqref{eq:potentiality.flat} and \eqref{eq:hatPsi.via.hatC}, and recalling $g(\mathsf{e},W)=0$, we furthermore have
	\begin{align*}
		g\left( (\nabla^M_Z\hat C)(X,Y),W \right)
		&= g(Z\star(X\star Y),W)
		= g(\hat C(Z,\hat C(X,Y)),W)
		\\
		&= g(\hat\Psi(Z,\hat\Psi(X,Y)),W)
		+\kappa^2r^2h(X,Y)h(Z,W)\,.
	\end{align*}
	Consequently, using that $\nabla^M_Z\mathsf{e}=-X$ due to~\eqref{eq:nabla.M.mixed},
	\begin{align*}
		\kappa r^2 h\left( (\nabla^N_Z\hat\Psi)(X,Y),W \right)
		&= g(\hat\Psi(Z,\hat\Psi(X,Y)),W)
		+\kappa^2r^2h(X,Y)h(Z,W)
		\\ &\quad
		+\kappa h(X,Y) g(Z,W)
		+\kappa h(Z,X)\,g(Y,W)
		+\kappa h(Z,Y)\,g(X,W)
		\\
		&= \kappa r^2 h\left( \hat\Psi(Z,\hat\Psi(X,Y)),W \right)
		+2\kappa^2 r^2 h(X,Y)h(Z,W)
		\\ &\qquad
		+\kappa^2 r^2 h(Z,X)\,h(Y,W)
		+\kappa^2 r^2 h(Z,Y)\,h(X,W)
	\end{align*}
	and hence, dividing through with $\kappa r^2\ne0$,
	\begin{align*}
		h\left( (\nabla^N_Z\hat\Psi)(X,Y),W \right)
		&= h\left( \hat\Psi(Z,\hat\Psi(X,Y)),W \right)
		\\ &\quad
		+\kappa \left( 2h(X,Y) h(Z,W)+h(Z,X)\,h(Y,W)+h(Z,Y)\,h(X,W)\right)\,.
	\end{align*}
	The claim follows.
\end{proof}

We end this section with a brief note on the potential of the abundant system $\mathcal S_N$, compared to its flat conification $\mathcal S_M$, using the terminology of Definition~\ref{defn:conification}. Proposition~\ref{prop:conification} guarantees the existence of a superintegrable potential $\hat V$ of $\mathcal S_M$.
Then, by \eqref{eq:structure.tensor},
\begin{equation}\label{eq:aux:upper.wilczynski}
	\nabla^M_B\partial_A\hat V
	= 3 \hat C_{AB}^J\partial_J\hat V
	- \frac{3}{n+1}g_{AB}\hat C^{IJ}_J\partial_I\hat V
	+ \frac{1}{n+1} g_{AB} \Delta_M\hat V\,,
\end{equation}
where $\hat C=\star$ and where $\nabla^M$ and $\Delta_M$ are the Levi-Civita and Laplace-Beltrami operator of $g$. We recall that in this section, capital Latin letters are used as indices that refer to adapted coordinates on $M$.
As before, we let $\hat\Psi=\ast$ and use $\nabla^N$ and $\Delta_N$ with the obvious meaning. Due to~\eqref{eq:nabla.M.horizontal}, with small Latin letters referring to the coordinates on $N$ and $\partial_0=\partial_r$, \eqref{eq:aux:upper.wilczynski} implies
\begin{equation*}
	\nabla^M_b\partial_a\hat V
	= \nabla^N_b\partial_a\hat V + r \kappa h_{ab}\partial_0\hat V
\end{equation*}
and, furthermore, we have
\begin{equation*}
	\hat C_{ab}^J\partial_J\hat V
	= \hat\Psi_{ab}^j\partial_j\hat V
	- r \kappa h_{ab}\partial_0\hat V
\end{equation*}
Taking the trace-free part of \eqref{eq:aux:upper.wilczynski} with respect to $h$,
\begin{equation*}
	\nabla^N_b\partial_a\hat V -\frac{1}{n}h_{ab}\Delta_N\hat V
	=
	3 \hat\Psi_{ab}^j\partial_j\hat V
	- \frac{3}{n}h_{ab}\hat\Psi_{i}^{ij}\partial_j\hat V\,.
\end{equation*}
Therefore, fixing $r=r_0$, $V=\hat V|_{r=r_0}$ is a function on $N$ and a solution of~\eqref{eq:Wilczynski}. Hence, $V$ is a potential for $\mathcal S_N$.
We leave a more thorough investigation of this relationship to future research. However, an example of two abundant systems related by conification is given in Section~\ref{sec:SW.and.sphere}.

\section{Examples}\label{sec:examples}

\subsection{Semi-simple Frobenius algebras}\label{sec:semisimple}

We consider a Frobenius algebra $(\mathbb K^n,\rho,\blackdiamond)$ with the standard \innerproduct\ $\rho$ and such that there is a basis $(v_i)_{1\leq i\leq n}$ with $v_i\blackdiamond v_j=\delta_{ij}v_i$. Here $\delta_{ij}$ denotes the Kronecker Delta. It follows, due to the Frobenius property, that
\[
	\rho(v_i,v_j)=\rho(v_i\blackdiamond v_i,v_j)=\rho(v_i,v_i\blackdiamond v_j)=\delta_{ij}\rho(v_i,v_i)\,.
\]
For simplicity, assume that $\mathbb K=\mathbb R$ and that $\rho$ is positive definite. By a change of coordinates, we thus achieve, for a rescaled orthonormal basis $(u_i)$, that
\begin{equation*}
	u_i\blackdiamond u_j=\mu_{i}\delta_{ij}u_i\,,\quad \rho(u_i,u_j)=\delta_{ij}.
\end{equation*}
for certain constants $\mu_i$.
Next, we consider $M=\mathbb R^n$ as an analytic manifold with the metric defined by $\rho$ in the canonical way. We identify $T_0M\simeq\mathbb R^n$ and we furthermore extend the basis $(u_i)$ to a frame $(e_i)$ of $TM$ in the parallel manner, i.e.\ such that $\nabla e_i=0$ for any value of~$i$, and $g(e_i,e_j)=\delta_{ij}$.
Then, assuming analyticity, \eqref{eq:potentiality.flat} determines a Hesse-Frobenius structure $\star$ on $(M,g)$. We obtain
\[
	(\nabla_lC_{ijk})_{x=0}=\rho(u_l\blackdiamond u_i\blackdiamond u_j,u_k)=\delta_{ij}\delta_{jk}\delta_{kl}\,\mu_i^2
\]
and so forth, and hence
\begin{equation*}
	e_i\star e_j=\lambda_i(x)\delta_{ij}e_i\,,
\end{equation*}
with functions $\lambda_i(x)$ satisfying the initial condition $\lambda_i(0)=\mu_i$, given by the Taylor series
\[
	\lambda_i(x) = \sum_{|\alpha|=0}^\infty \frac{\nabla_\alpha\lambda_i(0)}{\alpha!} x^\alpha
	=\sum_{k=0}^\infty \frac{k!\,\mu_i^{k+1}}{k!} x_i^k = \frac{\mu_i}{1-\mu_i x_i}
\]
where $\alpha$ is a multi-index and $|x_i\mu_i|<1$ (no summation over $i$).
If $\mu_i=0$, we obtain $\lambda_i(x)=0$. Otherwise,
\begin{equation}\label{eq:sesi.lambda_i}
	\lambda_i(x)=\frac{1}{c_i-x_i}
\end{equation}
for some constant $c_i=\frac{1}{\mu_i}$ ($\mu_i\ne0$).
Assume that \eqref{eq:sesi.lambda_i} holds, analogously, for all values of $i$. Then we obtain the commutative and associative product $\star$ with
\begin{equation}\label{eq:sesi.star}
	\partial_i\star \partial_j=\frac{\delta_{ij}}{c_i-x_i}\partial_i,
\end{equation}
which satisfies the Frobenius property with respect to the metric
\begin{equation}\label{eq:sesi.g}
	g_{\text{std}}=\sum_{k=1}^n dx_i\otimes dx_i.
\end{equation}
An analogous reasoning holds if $\mathbb K=\mathbb C$, leading to the same formulas \eqref{eq:sesi.star} and~\eqref{eq:sesi.g}.
Such a Frobenius manifold is called \emph{semi-simple}.
By a slight abuse of terminology, we shall call all solutions $(g_{\text{std}},C)$ with
\begin{equation}
	C = \sum_{k=m+1}^n\frac{1}{c_k-x_k}dx_k\otimes dx_k\otimes dx_k,\qquad
	0\leq m\leq n,
\end{equation}
\emph{semi-simple (in a degenerate sense)}, where $c_i\in\mathbb K$ are constants.
One straightforwardly verifies that they define flat Hesse-Frobenius manifolds.
We arrive at the following statement.
\begin{prop}\label{prop:sesi}
	A (maybe degenerately) semi-simple Frobenius algebra on $(\mathbb K^n,g_{\text{std}})$ corresponds to a second-order maximally superintegrable system with Hamiltonian
	\begin{equation}\label{eq:sesi}
		H=\sum_{i=1}^n p_ip_i + a_0\left( 4\sum_{j=1}^m x_j^2+\sum_{j=m+1}^nx_j^2\right)
		+ \sum_{j=1}^m a_jx_j+\sum_{j=m+1}^n \frac{a_j}{x_j^2} + a_{n+1}
	\end{equation}
	for some $m$ with $0\leq m\leq n$, up to coordinate changes. The $a_j\in\mathbb K$ are constant parameters.
\end{prop}
The (truly) semi-simple case is that when $m=0$.
\begin{proof}
	The Hamiltonian \eqref{eq:sesi} is obtained by carrying out steps 1 and 2 in the procedure detailed in Section~\ref{sec:HF2SIS} for the $(1,2)$-tensor field
	\[
		\hat C = \sum_{j=m+1}^n \frac{1}{c_j-x_j}\,dx_j\otimes dx_j\otimes \partial_{x_j}.
	\]
	The crucial step is the integration of \eqref{eq:Wilczynski.system}, using~\eqref{eq:structure.tensor}. Note that it is a system of partial differential equations of the form $\partial_iw=A_{i}w$, $i=1,\dots,n$, for $w=(V,\partial_1V,\dots,\partial_nV,\triangle V)^\mathsf{T}$, where $\triangle=\sum_{k=1}^n\partial_k^2$ is the usual Laplacian, and $A_i$ a suitable matrix. The integrability conditions for this system are satisfied generically, by our hypothesis, and thus the linear space of solutions is $(n+2)$-dimensional. One easily checks that the coefficient function of each of the parameters $a_j$ in \eqref{eq:sesi} is a solution, $j\in\{0,1,\dots,n+1\}$.	
	We compile the $n+2$ instances of $w$ of these solutions into the columns of a $(n+2)\times(n+2)$-matrix $\Omega$ and verify that its determinant is non-vanishing on its domain, $\det\Omega\propto\prod_{k=m+1}^n\frac{1}{x_j^3}\ne0$. Hence, $\Omega$ is a fundamental matrix for the system.
\end{proof}

\begin{rmk}
	We mention that the Hamiltonian \eqref{eq:sesi} is well-known in the literature, cf.\ \cite{KKM2018_book,KSV2023} for instance.
	Specifically, the Hamiltonians \eqref{eq:sesi} with $n=2$ and $m\in\{0,1,2\}$ correspond to the systems [E3], [E2], [E1] in \cite{KKPM2001}, respectively. For $n=3$, the corresponding superintegrable systems are labeled [I], [IV], [OO] and [O], cf.\ \cite{KKM2007_2Dflat,Capel_thesis}.
	
	Note that we include the dimension $n=2$ here. This is not a contradiction to the fact that the correspondence between Hesse-Frobenius manifolds and abundant systems on spaces of constant sectional curvature holds for $n\geq3$ only, cf.\ Section~\ref{sec:introduction}.
\end{rmk}

\subsection{A 4D degenerate semi-simple example}\label{sec:3+1}

As an example, consider the metric $g=dx_1^2+dx_2^2+dx_3^2+dx_4^2$ together with the cubic
\[
	C = -\frac{1}{x_1}dx_1^3 -\frac{1}{x_2}dx_2^3 -\frac{1}{x_3}dx_3^3,
\]
where $dx_k^3=dx_k\otimes dx_k\otimes dx_k$.
The structure tensor is then readily computed using \eqref{eq:structure.tensor}.
We then integrate \eqref{eq:Wilczynski} and obtain the potential
\begin{equation}\label{eq:3+1.example.potential}
	V(x) = a_0 \left( x_1^2 +x_2^2 +x_3^2 +4x_4^2 \right)
	+\frac{a_1}{x_1^2} +\frac{a_2}{x_2^2} +\frac{a_3}{x_3^2} +a_4x_4 +a_5\,,
\end{equation}
which leads to the Hamiltonian \eqref{eq:Hamilton} in the form
\begin{equation}\label{eq:3+1.example.Hamiltonian}
    H(x,p)=p^2
        +a_0\left( \sum_{j=1}^3 x_j^2+4x_4^2\right)
        +\sum_{j=1}^3\frac{a_j}{x_j^2}+a_4x_4+a_5\,,
\end{equation}
which we recognize as the Hamiltonian of the $4$-dimensional Smorodinski-Winternitz system with one degenerate direction.
The system's associated Killing tensor fields are given by \eqref{eq:shortcut}, or alternatively by \eqref{eq:shortcut.parametrised}.
In order to keep the notation short, we shall denote symmetric tensor products of $1$-forms in the condensed form
\begin{equation*}
	dx^i\,dx^j=\frac12\left( dx^i\otimes dx^j+dx^j\otimes dx^i\right),
\end{equation*}
since there is no risk of confusion.
The $10$-dimen\-sional linear space of solutions can then be represented by the basis $(K^{(k)})_{k\in\{1,\dots,10\}}$ as follows:
\begin{subequations}\label{eq:semisimple.KT}
	\begin{align}
		K^{(1)} &= dx_1^2\,, &
		K^{(2)} &= dx_2^2\,, &
		K^{(3)} &= dx_3^2\,, &
		K^{(4)} &= dx_4^2\,,
	\end{align}
	and
	\begin{align}
		K^{(5)} &= \left( x_2\,dx_1-x_1\,dx_2\right)^2\,, &
		K^{(6)} &= \left( x_3\,dx_2-x_2\,dx_3 \right)^2\,, \\
		K^{(7)} &= \left( x_3\,dx_1-x_1\,dx_3\right)^2\,, &
		K^{(8)} &= (x_4\,dx_3-x_3\,dx_4)dx_3\,, \\
		K^{(9)} &= (x_4\,dx_1-x_1\,dx_4)dx_1\,, &
		K^{(10)} &= (x_4\,dx_2-x_2\,dx_4)\,dx_2\,.
	\end{align}
\end{subequations}
Finally, we use \eqref{eq:preBD} to construct the integrals of the motion for the Hamiltonian \eqref{eq:3+1.example.Hamiltonian}. We recall that the integrability condition is satisfied automatically be the construction of $V$ and the compatible Killing tensor fields, as \eqref{eq:BD} is satisfied.
Explicitly, we obtain (omitting the obvious freedom of adding a arbitrary constant to each of the functions)
\begin{align*}
	F^{(1)} &= p_1^2 + a_0 x_1^2 +\frac{a_1}{x_1^2}
	\\
	F^{(2)} &= p_2^2 + a_0 x_2^2 + \frac{a_2}{x_2^2}
	\\
	F^{(3)} &= p_3^2 + a_0 x_3^2 + \frac{a_3}{x_3^2}
	\\
	F^{(4)} &= p_4^2 + 4a_0 x_4^2 + a_4x_4
	\\
	F^{(5)} &= (x_2p_1-x_1p_2)^2 + \frac{a_2 x_1^2}{x_2^2} +\frac{a_1 x_2^2}{x_1^2}
	\\
	F^{(6)} &= (x_3p_2-x_2p_3)^2 + \frac{a_3 x_2^2}{x_3^2} +\frac{a_2 x_3^2}{x_2^2}
	\\
	F^{(7)} &= (x_3p_1-x_1p_3)^2 + \frac{a_1 x_3^2}{x_1^2} +\frac{a_3 x_1^2}{x_3^2}
	\\
	F^{(8)} &= (x_4p_3-x_3p_4)p_3 -a_0 x_3^2x_4 +\frac{a_3x_4}{x_3^2} -\frac{a_4 x_3^2}{4}
	\\
	F^{(9)} &= (x_4p_3-x_3p_4)p_3 -a_0 x_1^2x_4 +\frac{a_1x_4}{x_1^2} -\frac{a_4 x_1^2}{4}
	\\
	F^{(10)} &= (x_4p_2-x_2p_4)p_2 -a_0 x_2^2x_4 +\frac{a_2x_4}{x_2^2} -\frac{a_4 x_2^2}{4}
\end{align*}
and observe $H=F^{(1)}+F^{(2)}+F^{(3)}+F^{(4)}$.

We comment that for systems in other dimensions, the Killing tensors are obtained analogously.

\subsection{Nilpotent Frobenius algebras}\label{sec:nilpotent}

We now consider algebras $(\mathbb K^n,\blackdiamond)$ of nilpotency class three (sometimes called $3$-step nilpotent). More precisely, we consider commutative and associative algebras such that any product of three elements is zero, $u\blackdiamond v\blackdiamond w=0$ $\forall u,v,w\in\mathbb K^n$. Hence, every element is nilpotent with index at most three.
We furthermore assume that there is a \innerproduct\ $\rho$ on $\mathbb K^n$ with respect to which $\blackdiamond$ satisfies the Frobenius property.
By the correspondence theorem in \cite{Vollmer2025_Frobenius}, it then follows that in a neighborhood $U$ around a point on $M=\mathbb K^n$, say $x=0$, there is a flat Hesse-Frobenius structure $\star$ with respect to the metric $g$ induced by the \innerproduct.

\subsubsection{First examples}\label{sec:shifted.truncated.polynomial.algebra}
We illustrate the procedure with a basic two-dimensional example, namely $\mathbb K^2$ with the \innerproduct
\[
 \rho = \begin{pmatrix} 0 & 1 \\ 1 & 0 \end{pmatrix}
\]
and together with the product defined by
\[
	c_{ijk}=\begin{cases}
		1 & \text{if $i=j=k=1$} \\
		0 & \text{otherwise} 
	\end{cases}.
\]
We introduce the structure constants $c_{ij}^k$ via $c_{ijk}=\rho_{ka}c_{ij}^a$ and denote the product by $\blackdiamond$, $e_i\blackdiamond e_j=c_{ij}^ke_k$ where $(e_l)_{l=1,2}$ is the canonical basis of $\mathbb K^2$.
Note that the product satisfies the nilpotency condition
\[
	\rho^{ab}c_{ija}c_{klb} = 0\,.
\]
As a consequence, the associativity condition is satisfied trivially.
\bigskip

The task at hand is now as follows:
consider the manifold $M=\mathbb K^2$ with the metric $g=2\,dx^1\,dx^2=dx^1\otimes dx^2+dx^2\otimes dx^1$ and a point $x\in M$. Note that $\rho=g_x$.
We seek a product of vector fields,
\[
	X\star Y:=C_{ijk}X^iY^jg^{kl}\partial_l\,,
\]
such that $\star_x=\blackdiamond$. One also quickly checks that, due to \eqref{eq:potentiality.flat},
\[
	(\nabla^m\star)_x = 0
\]
must be satisfied for all derivatives $m\geq1$. Assuming analyticity, this implies that $C_{ijk}$ must be constant in $\nabla$-affine coordinates, and hence $C_{ijk}=c_{ijk}$, i.e.
\[
	C=dx_1\otimes dx_1\otimes dx_1.
\]
We introduce $\hat C$ by raising one index using $g$,
\[
	\hat C=dx_1\otimes dx_1\otimes \partial_2,
\]
and we observe that $g^{ab}C_{abc}=0$.
The corresponding structure tensor is then obtained via~\eqref{eq:structure.tensor},
\[
	\hat T = 3 \hat C-0 = 3\,dx_1\otimes dx_1\otimes \partial_2\,.
\]
In order to obtain the potential, and thus the Hamiltonian, we consider~\eqref{eq:Wilczynski.system}, which is given by
\[
	\frac{\partial^2V}{\partial x_1^2}=3\frac{\partial V}{\partial x_2}\,,
	\qquad\qquad
	\frac{\partial^2V}{\partial x_2^2}=0\,,
\]
together with
\[
	\frac{\partial}{\partial x_1}\triangle V = 6\frac{\partial^2V}{\partial x_2^2}=0\,,
	\qquad\qquad
	\frac{\partial}{\partial x_2}\triangle V = 0\,,
\]
where $\triangle=2\frac{\partial^2}{\partial x_1\partial x_2}$.
Integrating this system, we find
\begin{equation}\label{eq:aux:shifted.2D.potential}
	V = a_1\,\left( x_1x_2 +\frac12\,x_1^3 \right)
	+ a_2\,\left( \frac{3x_1^2}{2}+x_2 \right)
	+ a_3\,x
	+ a_4
\end{equation}
with integration constants $a_1,a_2,a_3,a_4\in\mathbb K$.

We conclude this example by also solving~\eqref{eq:shortcut}.
Writing $K=K_{11}dx^2+2K_{12}dxdy+K_{22}dy^2$, this system becomes
\begin{align*}
	\frac{\partial K_{11}}{\partial x_1}&=0\,,
	&
	\frac{\partial K_{12}}{\partial x_1}&=-K_{22}\,,
	&
	\frac{\partial K_{22}}{\partial x_1}&=0\,,
	\\
	\frac{\partial K_{11}}{\partial x_2}&=2K_{22}\,,
	&
	\frac{\partial K_{12}}{\partial x_2}&=0\,,
	&
	\frac{\partial K_{22}}{\partial x_2}&=0\,.
\end{align*}
The solution hence is straightforwardly obtained as
\[
	K=(2b_1x_2+b_2)dx^2+2(-b_1x_1+b_3)dxdy+b_1dy^2\,,
\]
meaning that the associated Killing tensor fields are
\[
	K^{(1)}=2x_2\,dx_1^2-2x_1\,dx_1dx_2+dx_2^2\,,\qquad
	K^{(2)}=dx_1^2\,,\qquad
	K^{(3)}=dx_1dx_2\,.
\]
Finally, integrating \eqref{eq:preBD}, we obtain the integrals of motion corresponding to these Killing tensor fields:
\begin{align*}
	F^{(1)} &= 2(x_2p_2-x_1p_1)p_2+p_1^2
		-\frac{a_1}{8}(3x_1^2+2x_2)(x_1^2-2x_2)
		-a_2\,x_1(x_1^2-2x_2)
		-\frac{a_3}{2}\,(x_1^2-2x_2),
	\\
	F^{(2)} &= p_2^2
		+\frac{a_1}{2}x_1^2
		+a_2x_1,
	\\
	F^{(3)} &= p_1p_2
		+\frac{a_1}{2}\,x_1(x_1^2+2x_2)
		+\frac{a_2}{2}\,(3x_1^2+2x_2)
		+a_3x_1,
\end{align*}
where we omit the arbitrary constant that can be added to any integral.
\bigskip

We seek generalisations of this example in higher dimensions. We observe that the previous example can be viewed as the algebra $\mathcal R_m=\mathbb K[u]/\langle u^m\rangle$ with the standard addition, but with the multiplication $\star:\mathcal R_m\times\mathcal R_m\to\mathcal R_m$,
$$ u^i\star u^j=\begin{cases}
	u^{i+j+1} & \text{for $i+j<m-1$} \\
	0 & \text{otherwise}
\end{cases} , $$
(note that the power is increased by one compared to the usual multiplication).
Hence, every element of $\mathcal R_m$ is nilpotent of index \emph{at most} $m+1$.
The algebra $\mathcal R_m$ is also equipped with the \innerproduct\ $\rho$,
\begin{equation}
	\rho(u^k,u^l) =
	\begin{cases}
		1 & \text{if $k+l=m-1$} \\
		0 & \text{otherwise}
	\end{cases}.
\end{equation}
It is again easy to check that this defines a commutative, associative algebra that satisfies the Frobenius property.
The case $m=2$ corresponds to the earlier example, for which we have found the potential~\eqref{eq:aux:shifted.2D.potential}. Hence, we obtain the following Hamiltonian, using the coordinate names $x$ and $y$ for better legibility.
\begin{ex}
	The algebra $\mathcal R_2$ leads to the superintegrable potential
	\begin{equation}\label{eq:shifted.2D}
		V = a_1\,\left( xy + \frac12\,x^3 \right)
		+ a_2\,\left( \frac{3x^2}{2}+y \right)
		+ a_3\,x
		+ a_4
	\end{equation}
	and thus the Hamiltonian $H(x,y,p_1,p_2)=2p_1p_2+V(x,y)$, where $a_1,\dots,a_4\in\mathbb K$ are constant parameters.
	It is straightforward to confirm, using $\mathbb K=\mathbb C$ and the potential determined earlier, that this leads to the second-order superintegrable system commonly labeled E10 in the 2D classification of second-order superintegrable systems \cite{KKPM2001,KKM2007_2Dflat}.
\end{ex}

Analogously, we obtain the first non-trivial case, namely $m=3$.
\begin{ex}
	The algebra $\mathcal R_3$ leads to the superintegrable potential
	\begin{equation}\label{eq:shifted.3D}
		V = a_1\,\left( xz + \frac58 x^4 + \frac32 y x^2 + \frac12\,y^2 \right)
		+ a_2\,\left( 2x^3 + 3xy + z \right)
		+ a_3\,\left( \frac{3x^2}{2} + y \right)
		+ a_4\,x
		+ a_5
	\end{equation}
	and thus the Hamiltonian $H(x,y,z,p_1,p_2,p_3)=2p_1p_3+p_2^2+V(x,y,z)$, where $a_1,\dots,a_5\in\mathbb K$ are constant parameters.
	This is the system VII of the 3D classification of non-degenerate second-order superintegrable systems ($\mathbb K=\mathbb C$), cf.\ \cite{KKM2007_3Dflat,Capel_thesis}.
\end{ex}

\subsubsection{Truncated polynomial algebras}\label{sec:truncated.polynomial.algebra}

Inspired by our previous example, i.e.\ $\mathcal R_m$, we consider the univariate truncated polynomial algebras $\mathcal T_m:=\mathbb K[u]/\langle u^m\rangle$. One straightforwardly verifies that $\mathcal T_m$ is a Frobenius algebras equipped with the \innerproduct
\begin{equation}
	\rho(u^k,u^l) =
	\begin{cases}
		1 & \text{if $k+l=m-1$} \\
		0 & \text{otherwise}
	\end{cases}
\end{equation}
as in the case of $\mathcal R_m$.
This \innerproduct\ is represented, in the basis $\mathcal B=(1,u,u^2,\dots,u^{m-1})$ by the antidiagonal unit matrix
\begin{equation}
	\sigma|_{\mathcal B} = \begin{pmatrix}
		 & & 1 \\
		& \iddots & \\
		1 & & 
	\end{pmatrix}
\end{equation}
We call $\mathcal T_m=\mathbb K[x]/\langle x^m\rangle$ together with the standard addition, the standard (truncated) multiplication $\blackdiamond:\mathcal T_m\times\mathcal T_m\to\mathcal T_m$ and endowed with the \innerproduct\ $\rho$ the \emph{truncated polynomial algebra of length $m$}.
A direct computation verifies that $\mathcal T_m$ is a commutative and associative algebra satisfying the Frobenius property
$$ \rho(a\cdot b,c)=\rho(a,b\cdot c). $$

\begin{ex}
	The truncated polynomial algebra of length $m=2$ corresponds to the second-order superintegrable system commonly labeled \texttt{E8} in the 2D classification ($\mathbb K=\mathbb C$), see \cite{KKPM2001,KKM2007_2Dflat,Evans1990}.
	Indeed, consider the truncated polynomial algebra $\mathcal T_2\coloneq\mathbb K[x]/\langle x^2\rangle\simeq\mathbb K^2$ with the \innerproduct\ defined by
	\begin{equation*}
		\rho(1,1) = 0,\quad
		\rho(1,x)=\rho(x,1)=1,\quad
		\rho(x,x)=0.
	\end{equation*}
	The multiplication $\blackdiamond$ hence is encoded in the structure constants (with respect to the basis $(1,x)$)
	\begin{equation*}
		c_{11}^1=1\,,\quad
		c_{11}^2=0\,,\quad
		c_{12}^1=0\,,\quad
		c_{12}^2=1\,,\quad
		c_{22}^1=0\,,\quad
		c_{22}^2=0\,,	
	\end{equation*}
	which the remaining ones determined by commutativity.
	Hence,
	\begin{equation*}
		c_1=\begin{pmatrix} 1 & 0 \\ 0 & 1 \end{pmatrix}\,,\quad
		c_2=\begin{pmatrix} 0 & 1 \\ 0 & 0 \end{pmatrix}\,.
	\end{equation*}
	Substituting into our ansatz, we obtain, since $c_2c_2=0$,
	\begin{align*}
		C_{ij}^\ell
		&= \sum_{\mu=0}^\infty \sum_{\alpha_1+\alpha_2=\mu} \frac{\mu!}{\alpha_1!\alpha_2!} y_1^{\alpha_1}y_2^{\alpha_2}\,\,(c_1^{\alpha_1}c_2^{\alpha_2}\sigma_{ij})^\ell
		=
		\sum_{\mu=0}^\infty \sum_{\beta=0}^\mu \frac{\mu!}{\beta!(\mu-\beta)!} y_1^{\mu-\beta}\,y_2^{\beta}\,(c_2^{\beta}\sigma_{ij})^\ell
		\\
		&=
		\left( \sum_{\mu=0}^\infty y_1^{\mu}\,\sigma_{ij}^\ell \right)
		+ \left(
		\sum_{\mu=1}^\infty \mu\,\,y_1^{\mu-1}\,y_2\,(c_2\sigma_{ij})^\ell
		\right)
		=
		\frac{1}{1-y_1}\,\sigma_{ij}^\ell + \frac{y_2}{(1-y_1)^2}\,(c_1\sigma_{ij})^\ell
	\end{align*}
	We obtain the local $(1,2)$-tensor field
	\begin{equation*}
		\hat C=\frac{1}{1-x}\,(dx^2 \otimes \partial_x + ( dx\otimes dy+dy\otimes dx )\otimes\partial_y )
		+ \frac{y}{(1-x)^2}\,(dx^2\otimes\partial_y).	
	\end{equation*}
	Integration as explained earlier leads to the superintegrable potential
	\begin{equation}
		V_{\mathcal T_2}=a_1\,\frac{y}{(x-1)^3} + a_2\,(x-1)y + a_3\,\frac{1}{(x-1)^2} + a_4\,,
	\end{equation}
	where $a_1,\dots,a_4\in\mathbb K$ are constant parameters.
	The corresponding Hamiltonian is
	\begin{equation}\label{eq:truncated.length.2}
		H_{\mathcal T_2} = 2p_xp_y+V_{\mathcal T_2}\,.
	\end{equation}
	This is indeed the second-order superintegrable system commonly labeled \texttt{E8} (for $\mathbb K=\mathbb C$).
\end{ex}

\begin{ex}
	The truncated polynomial algebra of length $m=3$ corresponds to the $3$-dimensional non-degenerate second-order superintegrable system with potential
	\begin{equation}\label{eq:truncated.length.3}
		V_{\mathcal T_3} = a_1\left( (x-1)z + \frac12 y^2 \right)
			+ a_2\left( \frac{z}{(x-1)^3} - \frac{3y^2}{2(x-1)^4} \right)
			+ a_3\,\frac{y}{(x-1)^3}
			+ a_4\,\frac{1}{(x-1)^2}
			+ a_5\,,
	\end{equation}
	where $a_1,\dots,a_5\in\mathbb K$ are constant parameters,
	and the Hamiltonian
	\begin{equation}
		H_{\mathcal T_3} = 2p_xp_z+p_y^2+V_{\mathcal T_3}\,.
	\end{equation}
	This is the III of the 3D classification ($\mathbb K=\mathbb C$), see \cite{KKM2007_3Dflat,KKM2006_IV,KKM2005_III,Capel_thesis}.
\end{ex}

\subsection{Smorodinski-Winternitz system and generic system on the \texorpdfstring{$n$}{n}-sphere}\label{sec:SW.and.sphere}

As an example of the flat conification construction for abundant systems, see Section~\ref{sec:conification}, we consider the so-called ``generic system on the (complex) $n$-sphere'', whose name we are going to abbreviate the name to \emph{generic system} in this section. We find that its flat conification is the \emph{Smorodinski-Winternitz system} of dimension $n+1$.
To avoid singularities, we define the generic system on $S_+^n=\{(x_1,\dots,x_{n+1})\in\mathbb K_+^{n+1}\,\colon\ \lVert x\rVert=1\}$. We endow $S_+^n$ with the round metric obtained by pulling back the metric on $\mathbb K^{n+1}$ in the standard way. We write
\[
	h=\left[ \sum_{k=1}^{n+1} dx_k^2 \right]_{S_+^n}\,,
\]
for the metric on $S_+^n$, and take note that it is subject to $r^2=\sum_{k=1}^{n+1}x_k^2=1$. The potential of the generic system is given by
\[
	U=\sum_{k=1}^{n+1}\left.\frac{b_k}{x_k^2}\right|_{r^2=1}+b_0
\]
with $n+2$ parameters $b_k\in\mathbb K$, $k\in\{0,1,\dots,n+1\}$. It is a non-degenerate second-order superintegrable system, and indeed an abundant system.

On the other hand, we consider the Smorodinski-Winternitz system, restricted to $M=\mathbb K_+^{n+1}$, endowed with the metric
\[
	g=\sum_{k=1}^{n+1} dx_k^2 = dr^2+r^2 h\,,
\]
and with the superintegrable potential ($a_0,\dots,a_{n+1}\in\mathbb K$)
\[
	V=a_{n+2}\sum_{k=1}^{n+1}x_k^2+\sum_{k=1}^{n+1}\frac{a_k}{x_k^2}+a_0\,,
\]
which is one of the cases in Proposition~\ref{prop:sesi}.
Hence, also the Smorodinski-Winternitz system is an abundant system, cf.\ \cite{KKM2018_book,KSV2023}. The Smorodinski-Winternitz system also appears to be related to supersymmetric mechanics, cf.\ \cite{KKLNS2018}.

In line with our observation at the end of Section~\ref{sec:conification}, we now consider the restriction
\[
	V|_{r^2=1} = a_{n+2}\cdot 1 
	+\sum_{k=1}^{n+1}\left.\frac{a_k}{x_k^2}\right|_{r^2=1}+a_0\,,
\]
which yields a family of potentials for the generic system.
A quick inspection shows that the parameters $a_{n+2}$ and $a_0$ then parametrise the same freedom. Indeed, $U$ and $V|_{r^2=1}$ coincide if we identify $b_k=a_k$ for $k\in\{1,\dots,n\}$ and $b_0=a_0+a_{n+2}$.

This illustrates the flat conification of an abundant system on a space of non-zero constant sectional curvature $\kappa=1$.

\subsection{Associated Killing tensors for an example on a direct product}\label{sec:Killings}

We discuss the existence of Killing tensor fields in superintegrable systems of direct product form. Note that in this case, the system defined on the product automatically inherits $\frac{n_1(n_1+1)}{2}$ and $\frac{n_2(n_2+1)}{2}$ Killing tensors from the component systems, whose dimensions we denote by $n_1$ and $n_2$, respectively.
Hence, we directly obtain $ \frac12(n_1(n_1+1)+n_2(n_2+1))$ Killing tensor fields that are compatible with the superintegrable potentials constructed on the product.
This number is strictly smaller than $\frac12(n_1+n_2)(n_1+n_2+1)=\frac12(n_1(n_1+1)+n_2(n_2+1)+2n_1n_2)$, which is the number of linearly independent Killing tensor fields compatible with the superintegrable potential of the abundant system arising on the product space.
This means that the existence of an additional number of $n_1n_2$ independent Killing tensor fields is guaranteed by the direct product construction. We illustrate this with an example.

\subsubsection{An example in dimension four}\label{sec:2+2}

We consider a simple direct product of two copies of the example discussed in Section~\ref{sec:shifted.truncated.polynomial.algebra}.
Take the $4$-dimensional flat metric
\begin{equation*}
	g = \left(
	\begin{array}{rrrr}
		0 & 0 & 1 & 0 \\
		0 & 0 & 0 & 1 \\
		1 & 0 & 0 & 0 \\
		0 & 1 & 0 & 0
	\end{array}
	\right)
\end{equation*}
and the Frobenius pre-potential
\begin{equation*}
	\phi(x) = \frac16\left(x_1^3+x_2^3\right)\,,
\end{equation*}
which gives rise to the cubic tensor field
\begin{equation*}
	C_{ijk}=\delta_{ij}\delta_{ik}(\delta_{i1}+\delta_{i2}),
\end{equation*}
i.e.\ $C=\nabla^3\phi =d x_1\otimes d x_1\otimes d x_1+d x_2\otimes d x_2\otimes d x_2$.
It is straightforwardly checked that $C$ satisfies the conditions of a Hesse-Frobenius manifold.
By integration, one obtains the superintegrable potential
\begin{multline*}
	V = \left( \frac{x_1^3}{2} + x_1 x_3 + \frac{x_2(x_2^2 + 2x_4)}{2} \right)a_1
	+\left( \frac{3x_1^2}{2} + x_3 \right)a_2
	+\left( \frac{3x_2^2}{2} + x_4 \right)a_3
	+ a_4 x_1 +a_5 x_2 + a_6
\end{multline*}
and thus the Hamiltonian $H=2(p_1p_3+p_2p_4)+V$.
The potential $V$ is compatible with the following ten Killing tensors:
$6=3+3$ Killing tensor fields are directly inherited from the constituent systems, namely
\begin{equation}\label{eq:KTs.set.1}
\begin{aligned}
	K^{(1)} &= dx_1^2\,,
	&\quad
	K^{(2)} &= dx_1\,dx_3\,,
	&\quad
	K^{(3)} &= (x_3\,dx_1-x_1\,dx_3)\,dx_1+\frac12\,dx_3^2
\end{aligned}
\end{equation}
and, respectively,
\begin{equation}\label{eq:KTs.set.2}
	\begin{aligned}
	K^{(4)} &= dx_2^2\,,
	&\quad
	K^{(5)} &= dx_2\,dx_4\,,
	&\quad
	K^{(6)} &= (x_4\,dx_2-x_2\,dx_4)\,dx_2+\frac12\,dx_4^2\,.
\end{aligned}
\end{equation}
In addition, four new Killing tensor fields arise that mix the constituent systems:
\begin{equation}\label{eq:set.12}
\begin{aligned}
	K^{(7)} &= dx_1\,dx_2\,, \\
	K^{(8)} &= (x_2\,dx_1-x_1\,dx_2)\,dx_1+dx_2\,dx_3\,, \\
	K^{(9)} &= (x_1\,dx_2-x_2\,dx_1)\,dx_2+dx_1\,dx_4\,, \\
	K^{(10)} &= \left(x_4-\frac12x_2^2\right)dx_1^2 +\left(x_3-\frac12\,x_1^2\right)\,dx_2^2
	+(x_1\,dx_1-dx_3)(x_2\,dx_2-dx_4)\,.
\end{aligned}
\end{equation}
The system under consideration arises as a product of two copies of a 2D second-order superintegrable systems. Each copy admits three compatible Killing tensor fields, namely the Killing tensor fields $K^{(1)},K^{(2)},K^{(3)}$ and, respectively, $K^{(4)},K^{(5)},K^{(6)}$ are inherited from the constituent systems. In addition, the Killing tensor fields $K^{(7)},K^{(8)},K^{(9)}$ and $K^{(10)}$ arise in the product system. This agrees with the number of $n_1\cdot n_2=2\cdot 2=4$ additional Killing tensor fields, which we expect by our consideration in the beginning of this section.

We remark that the definition of a maximally superintegrable system in fact only requires $2n-1=7$ (for dimension $n=4$) Killing tensor fields, as long as their associated integrals of the motion are \emph{functionally} independent. However, it is proven in \cite{KSV2023} that the systems arising from the procedure in Section~\ref{sec:HF2SIS} always admit $\frac{n(n+1)}{2}=10$ \emph{linearly} independent second-order integrals of the motion, which are associated here to the Killing tensor fields $K^{(1)}$ to $K^{(10)}$. It is proven in \cite{KSV2023} that for a generic choice of the parameters $a_1$ to $a_6$ in the potential, there exists, in addition to the metric, a suitable choice of six Killing tensor fields from the $10$-dimensional space $\mathrm{span}(K^{(1)},\dots,K^{(10)})$ that lead to seven functionally independent integrals of the motion (including the Hamiltonian itself).

We conclude this example with a list of the integrals of the motion associated with this system, employing again \eqref{eq:preBD} to obtain the scalar part of each integral. We find, omitting again the obvious freedom of adding a arbitrary constant,
\begin{align*}
	F^{(1)} &= p_3^2 + \frac{a_1 x_1^2}{2} + a_2x_1
	\\
	F^{(2)} &= p_1p_3 +a_1\frac{x_1(x_1^2+2x_3)}{4} +a_2\,\frac{3x_1^2+2x_3}{4} +\frac{a_4 x_1}{2}
	\\
	F^{(3)} &= \frac12 p_1^2 +(x_3p_3-x_1p_1)p_3
	-a_1\,\frac{(3x_1^2+2x_3)(x_1^2-2x_3)}{16}
	-a_2\,\frac{x_1(x_1^2-2x_3)}{2}
	+a_4\,\frac{2x_3-x_1^2}{4}
\end{align*}
and analogous expressions for $F^{(4)}$, $F^{(5)}$ and $F^{(6)}$,
\begin{align*}
	F^{(4)} &= p_4^2 + \frac{a_1 x_2^2}{2} + a_3x_2
	\\
	F^{(5)} &= p_2p_4 +a_1\frac{x_2(x_2^2+2x_4)}{4} +a_3\,\frac{3x_2^2+2x_4}{4} +\frac{a_5 x_2}{2}
	\\
	F^{(6)} &= \frac12 p_2^2 +(x_4p_4-x_2p_2)p_4
	-a_1\,\frac{(3x_2^2+2x_4)(x_2^2-2x_4)}{16}
	-a_3\,\frac{x_2(x_2^2-2x_4)}{2}
	+a_5\,\frac{2x_4-x_2^2}{4}
\end{align*}
Furthermore,
\begin{align*}
	F^{(7)} &= p_3p_4 +a_1\frac{x_1x_2}{2} +a_3\,\frac{x_1}{2} +a_2\,\frac{x_2}{2}
	\\
	F^{(8)} &= (x_2p_3-x_1p_4)p_3 + p_1p_4
	+a_1\,\frac{(x_1^2+2x_3)x_2}{4}
	+a_2\,x_1x_2
	+a_3\,\frac{2x_3-x_1^2}{4}
	+a_4\,\frac{x_2}{2}
	\\
	F^{(9)} &= (x_1p_4-x_2p_3)p_4 + p_2p_3
	+a_1\,\frac{(x_2^2+2x_4)x_1}{4}
	+a_2\,\frac{2x_4-x_2^2}{4}
	+a_3\,x_1x_2
	+a_5\,\frac{x_1}{2}
	\\
	F^{(10)} &= \left(x_4-\frac12 x_2^2\right)p_3^2
	+\left(x_3-\frac12 x_1^2\right)p_4^2
	+ (x_1p_3-p_1)(x_2p_4-p_2)
	\\ &\quad
	+a_1\,\left(
			-\frac38 x_1^2x_2^2
			+ \frac14 x_1^2x_4
			+ \frac14 x_2^2x_3
			+ \frac12 x_3x_4
		\right)
	\\ &\quad
	-a_2\,\frac{x_1(x_2^2-2x_4)}{2}
	-a_3\,\frac{x_2(x_1^2-2x_3)}{2}
	+a_4\,\frac{2x_4-x_2^2}{4}
	+a_5\,\frac{2x_3-x_1^2}{4}.
\end{align*}
We observe that $H=2( F^{(2)}+F^{(5)} )$.

\subsubsection{An example in dimension eight}\label{sec:8D}

Note that the examples we have discussed earlier are also of the product form $(M,G,\divideontimes)$.
The product construction allows us to construct further examples from known ones, and particularly we are able to combine the examples already presented in order to obtain new systems on the respective product spaces. As an example, we combine the nilpotent example with the 3D Smorodinski-Winternitz system and the 1D Harmonic oscillator. We obtain the following 8D Hamiltonian system: its underlying metric is
\begin{equation*}
	G = \left(
	\begin{array}{rrrr rrrr}
		0 & 0 & 1 & 0 &&&& \\
		0 & 0 & 0 & 1 &&&& \\
		1 & 0 & 0 & 0 &&&& \\
		0 & 1 & 0 & 0 &&&& \\
		&&&& 1 & 0 & 0 & 0 \\
		&&&& 0 & 1 & 0 & 0 \\
		&&&& 0 & 0 & 1 & 0 \\
		&&&& 0 & 0 & 0 & 1
	\end{array}
	\right)
\end{equation*}
defined on a subset of $\mathbb K^{4+4}=\mathbb K^8$; note that the entries of $G$ that are not shown are supposed to vanish. The momenta $p_1,\dots,p_4$ correspond to $x$, the remaining momenta $p_5\,,\dots,p_8$ to $y$. The Hamiltonian of the product system thus takes the form 
\begin{equation*}
	H = 2p_1p_3+2p_2p_4+ p_5^2+p_6^2+p_7^2+p_8^2+V(x,y)\,,
\end{equation*}
where the potential is of the form
\begin{align*}
	V(x,y) &= a_1\left( \frac12x_1^3 +\frac12x_2^3 +x_1x_3 +x_2x_4 +\frac18x_5^2 +\frac18x_6^2 +\frac18x_7^2 +\frac12x_8^2 \right)
	\\
	&\qquad
	+ a_2\left( \frac{3x_1^2}{2}+x_3 \right)
	+ a_3\left( \frac{3x_2^2}{2}+x_4 \right)
	+ a_4x_1 +a_5x_2
	\\
	&\qquad\qquad
	+ a_6x_8 +\frac{a_7}{x_5^2} +\frac{a_8}{x_6^2} +\frac{a_9}{x_7^2} + a_{10}.
\end{align*}
This potential is compatible with $\frac12\cdot 8\cdot 9=36$ Killing tensor fields.
We are now going to list these Killing tensor fields.
First, we have the six pure squares of linear momenta
\[
dx_1^2\,,\quad dx_2^2\,,\quad\text{as well as}\quad dx_5^2\,,\quad dx_6^2\,,\quad dx_7^2\,,\quad dx_8^2\,,
\]
then the five mixed squares of linear
\[
dx_1\,dx_2\,,\quad
dx_1\,dx_3\,,\quad 
dx_2\,dx_4\,,
\]
and
\begin{equation}\label{eq:set.123.a}
	dx_1\,dx_8\,,\quad
	dx_2\,dx_8\,.
\end{equation}
Next, we have three symmetric products of the form
\begin{align*}
	& (x_8\,dx_5-x_5\,dx_8)dx_5\,, &
	& (x_8\,dx_6-x_6\,dx_8)dx_6\,, &
	& (x_8\,dx_7-x_7\,dx_8)dx_7
\end{align*}
and six of the form
\begin{subequations}\label{eq:set.123.b}
\begin{align}
	& (x_1\,dx_7-x_7\,dx_1)dx_7\,, &
	& (x_2\,dx_7-dx_2\,x_7)dx_7\,, &   
	& (x_2\,dx_5-x_5\,dx_2)dx_5\,, \\
	& (x_1\,dx_6-x_6\,dx_1)dx_6\,, &
	& (x_2\,dx_6-x_6\,dx_2)dx_6\,, &
	& (x_1\,dx_5-x_5\,dx_1)dx_5
\end{align}
\end{subequations}
and four combinations of the form
\begin{equation*}
	(x_1\,dx_2-x_2\,dx_1)dx_2+dx_1\,dx_4\,,\quad
	(x_2\,dx_1-x_1\,dx_2)dx_1+dx_2\,dx_3
\end{equation*}
and
\begin{equation}\label{eq:set.123.c}
	(x_8\,dx_2-x_2\,dx_8)dx_2+dx_4\,dx_8\,,\quad
	(x_8\,dx_1-x_1\,dx_8)dx_1+dx_3\,dx_8\,.
\end{equation}
We continue with two Killing tensor fields of the form
\[
(x_3\,dx_1-x_1\,dx_3)dx_1+\frac12\,dx_3^2\,,
\quad\text{and}\quad
(x_4\,dx_2-x_2\,dx_4)dx_2+\frac12\,dx_4^2\,,
\]
and three further ones, namely the pure squares of angular momenta
\begin{equation*}
	(x_7\,dx_6-x_6\,dx_7)^2 \,,\quad
	(x_6\,dx_5-x_5\,dx_6)^2 \,,\quad
	(x_7\,dx_5-x_5\,dx_7)^2 \,.
\end{equation*}
Finally, we have the six Killing tensor fields
\begin{subequations}\label{eq:set.123.d}
\begin{align}
	& (x_6\,dx_1-x_1\,dx_6)^2 +2(x_6\,dx_3-x_3\,dx_6)dx_6\,, \\
	& (x_7\,dx_1-x_1\,dx_7)^2 +2(x_7\,dx_3-x_3\,dx_7)dx_7\,, \\
	& (x_5\,dx_1-x_1\,dx_5)^2 +2(x_5\,dx_3-x_3\,dx_5)dx_5\,, \\
	& (x_6\,dx_2-x_2\,dx_6)^2 +2(x_6\,dx_4-x_4\,dx_6)dx_6\,, \\
	& (x_7\,dx_2-x_2\,dx_7)^2 +2(x_7\,dx_4-x_4\,dx_7)dx_7\,, \\
	& (x_5\,dx_2-x_2\,dx_5)^2 +2(x_5\,dx_4-x_4\,dx_5)dx_5
\end{align}
\end{subequations}
and eventually also the Killing tensor
\begin{equation*}
	\left(x_4-\frac12x_2^2\right)dx_1^2
	+\left(x_3-\frac12\,x_1^2\right)\,dx_2^2
	+(x_1\,dx_1-dx_3)(x_2\,dx_2-dx_4)\,.
\end{equation*}
Altogether, we have listed the
\[
6+5+9+4+2+3+6+1 = 36
\]
Killing tensor fields compatible with the $10$-parameter potential.
Out of these, $3+3+6+1=13$ are inherited from the two 2D, the 3D and the 1D constituent systems, respectively.
From the level of the two 4D constituents, an additional number of $4$ and $3$ Killing tensor fields, respectively, are inherited. Finally, in the 8D system, 16 additional Killing tensor fields arise.
This surplus of Killing tensor fields agrees with the expected number of $n_1\cdot n_2=4\cdot 4=16$ many, compare the discussion in the beginning of this section.
Note that their existence is automatically ensured through the direct product construction.
We leave it to the reader to determine the full integrals of motion associated to the system, which can be done by integrating \eqref{eq:preBD} for each $K^{(r)}$, recalling that the integrability condition \eqref{eq:BD} is satisfied by construction.

\section{Discussion}\label{sec:discussion}

In this paper, we have seen how abundant systems on manifolds of constant sectional curvature, in arbitrarily high dimension, can be systematically and geometrically constructed from Frobenius algebras based on the systems' underpinning Hesse-Frobenius structure. From these basic examples, further systems can be obtained on direct products, and systems on the $n$-sphere can be recovered from unital examples.
This demonstrates that already a small set of basic algebraic structures can efficiently be used to construct many second-order superintegrable systems.

While Hesse-Frobenius structures and abundant systems on spaces of constant sectional curvature only correspond 1-to-1 if the dimension is $n\geq3$, we may still obtain a subset of non-degenerate second-order superintegrable systems even in dimension two. Indeed, for $2$-dimensional superintegrable systems, we have seen that the systems labeled [E1], [E2], [E3], [E8] and [E10] in \cite{KKPM2001} arise from Hesse-Frobenius structures via the algebras considered here.

In contrast, in dimension three all non-degenerate second-order superintegrable systems on spaces of constant sectional curvature arise from Hesse-Frobenius structures. Indeed, it was proved in \cite{KKM2006_IV} that, in dimension $n=3$, non-degenerate second-order superintegrable systems are necessarily abundant systems.
Let us therefore confront the examples discussed in this paper with the known classification of flat non-degenerate 3-dimensional superintegrable systems, cf.\ \cite{Capel_thesis,KKM2006_IV,KKM2007_3Dflat}. These references work over the complex numbers, $\mathbb K=\mathbb C$. We use the labels coined in these references. Moreover, we indicate the underlying $3$-dimensional algebras in the classification \cite{KSTS2021}, together with those from \cite{FP2009} as determined in \cite{KSTS2021}. For a further comparison, e.g.\ with classical results by Peirce \cite{Peirce1881}, Scheffers \cite{Scheffers1891} and Study \cite{Study1890}, see \cite{KSTS2021} and the references therein. We find (the labels on the left refer to the superintegrable system following the nomenclature of \cite{KKM2007_3Dflat,Capel_thesis}):

\begin{description}[labelwidth=.8cm,leftmargin=.8cm,labelsep=0cm,align=left,itemsep=1pt]
	\item[I] The `pure' Smorodinski-Winternitz system. This is a direct product of three $1$-dimensional semi-simple components. Compare Proposition~\ref{prop:sesi}, specifically \eqref{eq:sesi} with $m=0,n=3$.
	The underlying algebra is labeled $U^3_2$ in \cite{KSTS2021} and $d_{14}$ in \cite{FP2009}.
	\item[II] A direct product of the $2$-dimensional truncated polynomial algebra, see \eqref{eq:truncated.length.2}, together with a $1$-dimensional semi-simple component, cf.\ \eqref{eq:sesi} with $m=0,n=1$.
	The underlying algebra is labeled $U^3_3$ in~\cite{KSTS2021} and $d_{12}$ in \cite{FP2009}.
	\item[III] The $3$-dimensional truncated polynomial algebra, see \eqref{eq:truncated.length.3}.
	The underlying algebra is labeled $U^3_4$ in~\cite{KSTS2021} and $d_{19}$ in \cite{FP2009}.
	\item[IV] A direct product of two $1$-dimensional semi-simple components and a $1$-dimensional trivial component, see Proposition~\ref{prop:sesi}, specifically \eqref{eq:sesi} with $m=1,n=3$.
	The underlying algebra is labeled $S^3_3$ in \cite{KSTS2021} and $d_{6}$ in \cite{FP2009}.
	\item[V] A direct product of the truncated polynomial algebra, cf.\ \eqref{eq:truncated.length.2}, and a $1$-dimensional trivial component.
	The underlying algebra is labeled $S^3_4$ in \cite{KSTS2021} and $d_{5}$ in \cite{FP2009}.
	\item[VI] A direct product of a $2$-dimensional shifted truncated algebra, cf.\ \eqref{eq:shifted.2D}, and a $1$-dimensional semi-simple component.
	The underlying algebra is labeled $S^3_2$ in \cite{KSTS2021} and $d_{2}$ in \cite{FP2009}.
	\item[VII] The $3$-dimensional shifted truncated algebra, cf.\ \eqref{eq:shifted.3D}.
	The underlying algebra is labeled $S^3_1$ in~\cite{KSTS2021} and $d_{16}$ in \cite{FP2009}.
	\item[O] The Harmonic oscillator. This is the $3$-dimensional trivial algebra. Compare Proposition~\ref{prop:sesi}, specifically \eqref{eq:sesi} with $m=3,n=3$.
	The underlying algebra is labeled $C^3_0$ in \cite{KSTS2021}. It is not targeted in \cite{FP2009}.
	\item[OO] A direct product of the $2$-dimensional trivial system and a $1$-dimensional semi-simple component. Compare Proposition~\ref{prop:sesi}, specifically \eqref{eq:sesi} with $m=2,n=3$.
	The underlying algebra is labeled $W^3_4$ in \cite{KSTS2021} and $d_{1}$ in \cite{FP2009}.
	\item[A] A direct product of the $2$-dimensional shifted truncated algebra, cf.\ \eqref{eq:shifted.2D}, and a $1$-dimensional trivial component.
	The underlying algebra is labeled $W^3_1$ in \cite{KSTS2021} and $d_{15}$ in \cite{FP2009}.
\end{description}
In addition, we mention the $3$-dimensional superintegrable system \textbf{VIII} on the complex $3$-sphere. It is the `complex generic system on the $3$-sphere', whose flat conification is the complex $4$-dimensional Smorodinski-Winternitz system, analogous to Section~\ref{sec:SW.and.sphere}. The system VIII is therefore a restriction, in the sense of converse conification, of the $4$-dimensional system arising locally from the semi-simple algebra \eqref{eq:sesi} with $m=0,n=4$.

Without going further into details (and for $\mathbb K=\mathbb C$), we remark that up to so-called \emph{Stäckel transformations}, i.e.\ conformal rescalings of the Hamiltonian that preserve superintegrability, all \emph{conformally flat} $3$-dimensional non-degenerate second-order superintegrable systems are given by the systems I, II, III (or V), IV, VI, VII, O, OO, A and VIII, see \cite{Capel_thesis}.
Informally speaking, we may therefore say that the Frobenius algebras discussed here -- $\mathcal T_m$, $\mathcal R_m$ and (degenerately) semi-simple ones -- locally represent all conformally flat 3D non-degenerate second-order superintegrable systems up to conification and Stäckel transformation.
	
It is worth remarking that our algebraic characterisation of the systems I---A is remarkably clean and concise. In particular, it appears to be significantly more succinct than that given in \cite{KKM2007_3Dflat}. This is also true in comparison to~\cite{Capel_thesis}, where, however, a viewpoint of conformal (Stäckel) equivalence classes is adopted. In turn, this appears to imply that the algebras under III and V in the above list ($U^3_4$ and $S^3_4$ \cite{KSTS2021} or $d_{19}$ and $d_5$ \cite{FP2009}, respectively) are related by some ``algebraic shadow'' of Stäckel equivalence. Exploring this perspective is left to future research.

On a similar note, we close with a remark on deformations.
Figure~5.1 of~\cite{Capel_thesis} shows a subideal containment diagram that represents a hierarchy of non-degenerate second-order superintegrable systems based on Stäckel transformations (i.e.\ conformal rescalings) and contractions. In Figure~1 of \cite{FP2009}, on the other hand, the moduli space of 3-dimensional associative algebras is depicted as a graph that illustrates versal deformations of such algebras. The resulting hierarchy, restricted to the algebras considered here, appears identical to that obtained from \cite{Capel_thesis}. We leave it to future research to investigate this correspondence more thoroughly.

\section*{Acknowledgments}

The author is grateful to A.~Bolsinov, J.~Kress, V.~Matveev and K. Schöbel for discussions, and acknowledges support by the SFB/CRC 1624 \emph{Higher structures, moduli spaces and integrability} at the University of Hamburg, by the \emph{Forschungsfonds} of the Department of Mathematics at the University of Hamburg, as well as by the University of New South Wales Sydney and by the Matrix Institute in Creswick, Australia.

{\small\printbibliography}

\end{document}